\documentclass[preprints,review,accept,oneauthor,pdftex]{mdpi} 
\firstpage{1}
\pubvolume{xx}
\issuenum{1}
\articlenumber{5}
\pubyear{2019}
\copyrightyear{2019}
\history{Received: date; Accepted: date; Published: date}
\pdfoutput=1
\usepackage{amssymb}
\numberwithin{equation}{section}

\Title{Effective field theories}
\Author{Andrey Grozin $^{1,2}$}
\AuthorNames{Andrey Grozin}
\address{%
$^{1}$ \quad Budker Institute of Nuclear Physics, Novosibirsk 630090, Russia\\
$^{2}$ \quad Novosibirsk State University, Novosibirsk 630090, Russia}
\corres{Correspondence: A.G.Grozin@inp.nsk.su}
\abstract{A pedagogical introduction to low-energy effective field theories.
In some of them, heavy particles are ``integrated out''
(a typical example --- the Heisenberg--Euler EFT);
in some heavy particles remain but some of their degrees of freedom are ``integrated out''
(Bloch--Nordsieck EFT).
A large part of these lectures is, technically, in the framework of QED.
QCD examples, namely, decoupling of heavy flavors and HQET,
are discussed only briefly.
However, effective field theories of QCD are very similar to the QED case,
there are just some small technical complications:
more diagrams, color factors, etc.
The method of regions provides an alternative view at low-energy effective theories;
it is also briefly introduced.}
\begin{document}
\section{Introduction}
\label{S:Intro}

We don't know \emph{all} physics up to \emph{infinitely high} energies
(or down to \emph{infinitely small} distances).
Therefore, \emph{all} our theories are effective low-energy
(or large-distance) theories
(except \emph{The Theory of Everything} if such a thing exists).

There is a high energy scale $M$ (and a short distance scale $1/M$)
where an effective theory breaks down.
We want to describe light particles (with masses $m_i\ll M$)
and their interactions at low energies,
i.\,e., with characteristic momenta $p_i\ll M$
(or, equivalently, at large distances $\gg1/M$).
To this end, we construct an effective Lagrangian containing the light fields.
Physics at small distances $\lesssim1/M$ produces
local interactions of these fields.
The Lagrangian contains all possible operators
(allowed by symmetries of our theory).
Coefficients of operators of dimension $n+4$ are proportional to $1/M^n$.
If $M$ is much larger than energies we are interested in,
we can retain only renormalizable terms (dimension 4),
and, perhaps, a power correction or two.

For more information about effective field theories see the textbook~\cite{PB:16}.

\section{Heisenberg--Euler EFT, heavy flavors in QCD}
\label{S:EFT}

A low-energy effective Lagrangian has been first constructed~\cite{EK:35}
to describe photon--photon scattering at $\omega \ll M$.
It contained dimension 8 operators made of four $F_{\mu\nu}$.
Later an effective Lagrangian for arbitrarily strong
homogeneous electromagnetic field has been constructed~\cite{HE:36}.
It contained all powers of $F_{\mu\nu}$ but no terms with derivatives.
It is not quite EFT in the modern sense:
for each dimensionality not all possible operators are included.
In QCD with a heavy flavor with mass $M$,
processes with characteristic energies $\ll M$ can be described
by an effective Lagrangian without this flavor.

\subsection{Photonia}
\label{S:Photonia}

\begin{figure}[ht]
\begin{center}
\begin{picture}(94,46)
\put(21,25){\makebox(0,0){\includegraphics{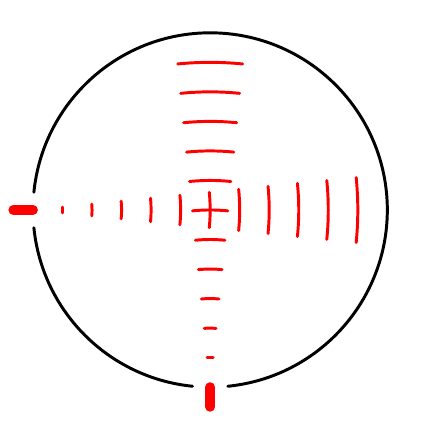}}}
\put(73.5,25){\makebox(0,0){\includegraphics{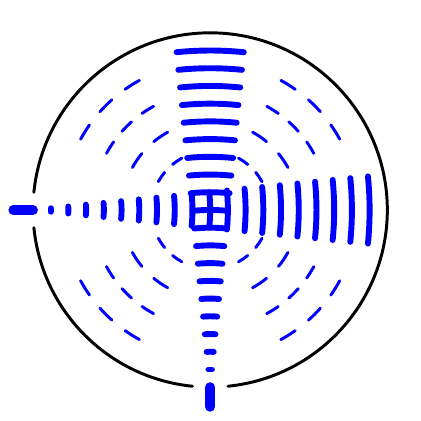}}}
\put(21,0){\makebox(0,0)[b]{a}}
\put(73,0){\makebox(0,0)[b]{b}}
\end{picture}
\end{center}
\caption{Scattering of low-energy photons}
\label{F:Photonia}
\end{figure}

Let's imagine a country, Photonia,
in which physicists have high-intensity sources
and excellent detectors of low-energy photons,
but they don't have electrons
and don't know that such a particle exists%
\footnote{We indignantly refuse to discuss the question
``What the experimentalists and their apparata are made of?''
as irrelevant.}.
At first their experiments (Fig.~\ref{F:Photonia}a)
show that photons do not interact with each other.
They construct a theory, Quantum PhotoDynamics (QPD),
with the Lagrangian
\begin{equation}
L_0 = - \frac{1}{4} F_{\mu\nu} F^{\mu\nu}\,.
\label{Photonia:L0}
\end{equation}
But later, after they increased the luminosity (and energy)
of their ``photon colliders'' and the sensitivity of their detectors,
they discover that photons do scatter,
though with a very small cross-section (Fig.~\ref{F:Photonia}b).
They need to add some interaction terms to this Lagrangian.

There are no dimension 6 gauge-invariant operators, because
\begin{equation}
F_\lambda{}^\mu F_\mu{}^\nu F_\nu{}^\lambda = 0
\label{Photonia:O6}
\end{equation}
(this algebraic fact reflects $C$-parity conservation).
All operators with derivatives reduce to
\begin{equation}
O = \bigl( \partial^\mu F_{\mu\lambda} \bigr) \bigl( \partial_\nu F^{\nu\lambda} \bigr)
\label{Photonia:O}
\end{equation}
plus full derivatives;
this operator vanishes due to equations of motion $\partial_\mu F^{\mu\nu} = j^\nu = 0$.
On-shell matrix elements of such operators vanish;
therefore, we may omit them from the Lagrangian
without affecting the $S$-matrix.

Interaction operators first appear at dimension 8:
\begin{equation}
O_1 = \left(F_{\mu\nu} F^{\mu\nu}\right)^2\,,\qquad
O_2 = F_{\mu\nu} F^{\nu\alpha} F_{\alpha\beta} F^{\beta\mu}\,.
\label{Photonia:O8}
\end{equation}
Hence the QPD Lagrangian which incorporates the photon--photon interaction is
\begin{equation}
L = L_0 + L_1\,,\qquad
L_1 = c_1 O_1 + c_2 O_2\,,
\label{Photonia:L1}
\end{equation}
where the coefficients $c_{1,2}\sim1/M^4$,
$M$ is some large mass (the scale of new physics).
Of course, operators of dimensions $>8$ can be also included,
multiplied by higher powers of $1/M$,
but their effect at low energies is much smaller.
Physicists from Photonia can extract the two parameters $c_{1,2}$
from two experimental results,
and predict results of infinitely many measurements.

We are working at the order $1/M^4$;
therefore, the photon--photon interaction vertex
can appear in any diagram at most once.
The only photon self-energy diagram vanishes:
\begin{equation}
\raisebox{-12mm}{\includegraphics{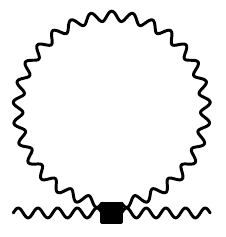}} = 0\,.
\label{Photonia:Pi0}
\end{equation}
Therefore, the full photon propagator is equal to the free one.
There are no loop corrections to the 4-photon vertex,
and hence the interaction operators~(\ref{Photonia:O8})
don't renormalize.

\subsection{Qedland}
\label{S:Qedland}

In the neighboring country Qedland physicists are more advanced.
In addition to photons, they know electrons and positrons,
and investigate their interactions at energies $E\sim M$
($M$ is the electron mass).
They have constructed a nice theory, QED, which describe
their experimental results%
\footnote{They don't know muons, but this is another story.}.

Physicists from Qedland understand that QPD constructed in Photonia
is just a low-energy approximation to QED.
The coefficients $c_{1,2}$ can be calculated by matching.
We calculate the amplitude of photon--photon scattering
at low energies in the full theory (QED),
expanded in the external momenta up to the 4-th order,
and equate it to the same amplitude in the effective theory (QPD):
\begin{equation}
\raisebox{-10.5mm}{\includegraphics{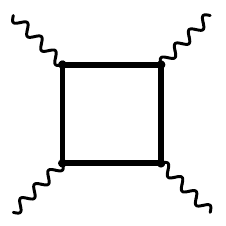}} =
\raisebox{-5.5mm}{\includegraphics{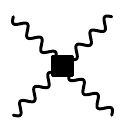}}\,.
\label{Qedland:Matching}
\end{equation}
After expanding the QED diagrams in the external momenta
they reduce to the vacuum integrals
\begin{equation}
\raisebox{-6.5mm}{\begin{picture}(14,17)
\put(7,7){\makebox(0,0){\includegraphics{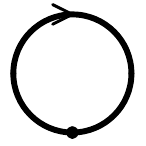}}}
\put(7,17){\makebox(0,0)[t]{$k$}}
\put(7,10){\makebox(0,0)[b]{$n$}}
\end{picture}} =
\frac{1}{i \pi^{d/2}} \int \frac{d^d k}{D^n} = M^{d-2n} V(n)\,,\quad
D = M^2-k^2-i0\,,\quad
V(n) = \frac{\Gamma\left(n-\frac{d}{2}\right)}{\Gamma(n)}\,.
\label{Qedland:V1}
\end{equation}
This QED scattering amplitude is finite at $\varepsilon \to 0$.
To reproduce it, the interaction term in the QPD Lagrangian~(\ref{Photonia:L1})
must be~\cite{EK:35}
\begin{equation}
L_1 = \frac{\pi \alpha^2}{180 M^4}
\left( - 5 O_1 + 14 O_2 \right)\,.
\label{Qedland:L1}
\end{equation}

It is not difficult to calculate the two-loop correction
to the Lagrangian~(\ref{Qedland:L1}).
The two-loop scattering amplitude in QED
reduces to the two-loop vacuum integrals
\begin{equation}
\begin{split}
&\raisebox{-13.2mm}{\begin{picture}(22,28)
\put(11,14){\makebox(0,0){\includegraphics{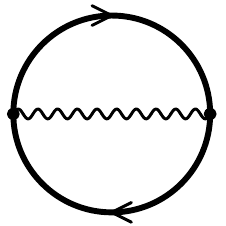}}}
\put(11,28){\makebox(0,0)[t]{$k_1$}}
\put(11,0){\makebox(0,0)[b]{$k_2$}}
\put(11,22){\makebox(0,0){$n_1$}}
\put(11,6){\makebox(0,0){$n_2$}}
\put(11,17){\makebox(0,0){$k_1-k_2$}}
\put(11,11){\makebox(0,0){$n_3$}}
\end{picture}} =
\frac{1}{(i\pi^{d/2})^2}
\int \frac{d^d k_1\,d^d k_2}{D_1^{n_1} D_2^{n_2} D_3^{n_3}}
= M^{2(d-n_1-n_2-n_3)} V(n_1,n_2,n_3)\,,\\
&D_1 = M^2 - k_1^2\,,\quad
D_2 = M^2 - k_2^2\,,\quad
D_3 = - (k_1-k_2)^2\,,\\
&V(n_1,n_2,n_3) =
\frac{\Gamma\left(\frac{d}{2}-n_3\right)
\Gamma\left(n_1+n_3-\frac{d}{2}\right)\Gamma\left(n_2+n_3-\frac{d}{2}\right)
\Gamma(n_1+n_2+n_3-d)}%
{\Gamma\left(\frac{d}{2}\right)\Gamma(n_1)\Gamma(n_2)\Gamma(n_1+n_2+2n_3-d)}
\end{split}
\label{Qedland:V2}
\end{equation}
($-i0$ is assumed in all denominators).

\subsection{Coulomb potential}
\label{S:Coulomb}

Let's suppose that physicists in Photonia have some classical
(infinitely heavy) charged particles,
and can manipulate them at their will.
If a particle with charge $-e$ moves along a world line $l$,
the action contains the interaction term
\begin{equation}
S_{\text{int}} = e \int_l d x^\mu A_\mu(x)
\label{Coulomb:S}
\end{equation}
in addition to the photon field action.
The integrand $\exp(iS)$ in the Feynman path integral
contains a phase factor
\begin{equation}
W_l = \exp \left(i e \int_l d x^\mu A_\mu(x) \right)
\label{Coulomb:W}
\end{equation}
called the Wilson line.
The vacuum-to-vacuum transition amplitude in the presence
of classical charges is thus the vacuum average
of the corresponding Wilson lines%
\footnote{In Sect.~\ref{S:Feyn} we'll see that the propagator of a heavy charged particle
in the effective theory which describes its interaction with soft photons
is the straight Wilson line.}.

Suppose two charges $e$ and $-e$ stay at rest
at some distance $\vec{r}$ during some (large) time $T$.
The energy of this system is $U(\vec{r})$ ---
the interaction potential of the charges.
The vacuum transition amplitude is
\begin{equation}
\raisebox{-15mm}{\begin{picture}(20,33)
\put(10,16.5){\makebox(0,0){\includegraphics{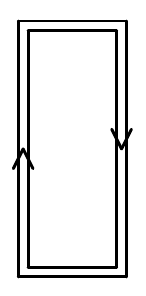}}}
\put(3,2){\makebox(0,0){$0$}}
\put(17,2){\makebox(0,0){$\vec{r}$}}
\put(3,31){\makebox(0,0){$T$}}
\end{picture}} =
e^{-i U(\vec{r}) T}
\label{Coulomb:vac}
\end{equation}
(we don't care what happens near its lower and upper ends
because $T \gg r$).

The zeroth-order term in the vacuum average of any Wilson loop is 1.
It is convenient (though not necessary) to use the Coulomb gauge
to calculate the first correction.
In this gauge, there is Coulomb photon with the propagator
\begin{equation}
D^{00}(q) = - \frac{1}{\vec{q}^{\,2}}
\label{Coulomb:D00}
\end{equation}
(it propagates instantaneously)
and transverse photon with the propagator
\begin{equation}
D^{ij}(q) = \frac{1}{q^2+i0}
\left(\delta^{ij}-\frac{q^i q^j}{\vec{q}^{\,2}}\right)\,.
\label{Coulomb:Dij}
\end{equation}
Wilson lines along the 0 direction only interact with Coulomb photons.
The self-energy of the classical particle vanishes:
\begin{equation}
\raisebox{-15.2mm}{\includegraphics{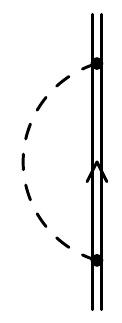}} = 0\,,
\label{Coulomb:self}
\end{equation}
because the particle propagates along time,
and the Coulomb photon along space.

Therefore, there is just one contribution at the order $e^2$:
\begin{equation}
\raisebox{-15.7mm}{\begin{picture}(24,33)
\put(10,16.5){\makebox(0,0){\includegraphics{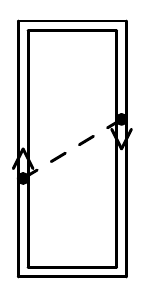}}}
\put(3,2){\makebox(0,0){$0$}}
\put(17,2){\makebox(0,0){$\vec{r}$}}
\put(3,31){\makebox(0,0){$T$}}
\put(2.5,13.5){\makebox(0,0){$\tau$}}
\put(21.5,19.5){\makebox(0,0){$\tau+t$}}
\end{picture}}
= - i\,e^2\,T\,\int D^{00}(t,\vec{r})\,dt
= - i\,e^2\,T\,\int \frac{d^{d-1}\vec{q}}{(2\pi)^{d-1}}\,
D^{00}(0,\vec{q})\,e^{i\,\vec{q}\cdot\vec{r}}
\label{Coulomb:W1}
\end{equation}
(integration in $\tau$ gives $T$).
Comparing it with $1 - i U(\vec{r}) T$~(\ref{Coulomb:vac}),
we obtain the Fourier transform of the potential
\begin{equation}
U(\vec{q}) = e^2 D^{00}(0,\vec{q}) = - \frac{e^2}{\vec{q}^{\,2}}\,;
\label{Coulomb:Uq}
\end{equation}
at $d=4$ the Coulomb potential is
\begin{equation}
U(\vec{r}) = - \frac{\alpha}{r}\,.
\label{Coulomb:U}
\end{equation}

What about corrections?
Vertex corrections vanish for the same reason as~(\ref{Coulomb:self});
crossed-box diagrams vanish because Coulomb photons propagate instantaneously,
and the time orderings of the vertices on the left line
and on the right one cannot be opposite:
\begin{equation}
\raisebox{-15.2mm}{\begin{picture}(22,32)
\put(11,16){\makebox(0,0){\includegraphics{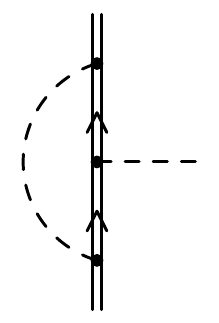}}}
\end{picture}} = 0\,,\qquad
\raisebox{-15.2mm}{\begin{picture}(22,32)
\put(11,16){\makebox(0,0){\includegraphics{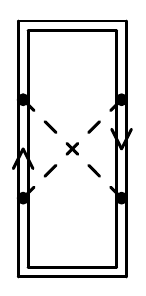}}}
\end{picture}} = 0\,.
\end{equation}
We don't need two-particle-reducible diagrams like
\begin{equation}
\raisebox{-15.2mm}{\begin{picture}(22,32)
\put(11,16){\makebox(0,0){\includegraphics{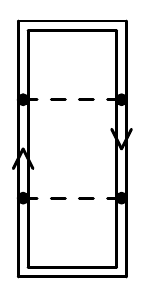}}}
\end{picture}}
\end{equation}
because they match higher orders of expansion
of the exponent~(\ref{Coulomb:vac}).
Only corrections to the photon propagator can contribute.
But there are no such corrections in QPD.
Hence the Coulomb potential~(\ref{Coulomb:U})
is exact in this theory.

In the presence of sources of the photon field,
the dimension 6 operator $O$~(\ref{Photonia:O})
cannot be ignored.
The QPD Lagrangian now contains an extra term
\begin{equation}
L_c = c O\,,
\label{Coulomb:Lc}
\end{equation}
where $c\sim1/M^2$ by dimensionality.
This term produces the contribution to the photon self-energy
\begin{equation}
\raisebox{-5mm}{\begin{picture}(26,9.5)
\put(13,4.75){\makebox(0,0){\includegraphics{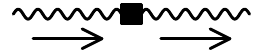}}}
\put(6.5,0){\makebox(0,0)[b]{$q$}}
\put(19.5,0){\makebox(0,0)[b]{$q$}}
\put(1,9.5){\makebox(0,0)[t]{$\mu$}}
\put(25,9.5){\makebox(0,0)[t]{$\nu$}}
\end{picture}}
= 2 i c q^2 \left(q^2 g_{\mu\nu} - q_\mu q_\nu\right)\,.
\label{Coulomb:Pic}
\end{equation}
The aim of an effective theory is to reproduce the $S$-matrix,
so, we may neglect operators vanishing due to equations of motion
(in particular, the self-energy~(\ref{Coulomb:Pic}) vanishes
at the photon mass shell $q^2=0$,
but is needed in a virtual photon line
exchanged between classical sources).
Therefore, the term~(\ref{Coulomb:Lc}) reduces to $c j_\mu j^\mu$,
where $j^\mu$ is the external (classical) current.
For the classical charges $e$ and $-e$,
it leads to the contact interaction%
\footnote{It is difficult to observe a $\delta$-function potential
in the interaction of classical charged particles.
But this interaction is essential if the particles
are quantum-mechanical --- it shifts energies of $S$-wave states.}
\begin{equation}
U_c(\vec{r}) = 2 c \delta(\vec{r})\,.
\label{Coulomb:Uc}
\end{equation}

What can physicists in Qedland say about the interaction potential
between classical charged particles?
In QED the photon self-energy is gauge invariant
because there are no off-shell charged external lines.
We  can use the covariant-gauge self-energy $\Pi(q^2)$ in the Coulomb-gauge propagator:
\begin{equation}
D^{00}(0,\vec{q}) = - \frac{1}{\vec{q}^{\,2}} \frac{1}{1 - \Pi(-\vec{q}^{\,2})}\,,\quad
U(\vec{q}) = e_0^2 D^{00}(0,\vec{q})\,.
\label{Coulomb:QED}
\end{equation}
In macroscopic measurements the potential at $\vec{q} \to 0$ is obtained:
\begin{equation}
U(\vec{q}) \to - \frac{e_0^2}{\vec{q}^2} \frac{1}{1 - \Pi(0)}
= - \frac{e_{\text{os}}^2}{\vec{q}^2}\,.
\label{Coulomb:macro}
\end{equation}
Here $e_{\text{os}}$ is the charge in the on-shell renormalization scheme.
The $\mathcal{O}(q^2)$ term in $\Pi(q^2)$ determines the contact interaction
constant $c$~(\ref{Coulomb:Uc}).

In the on-shell renormalization scheme
\begin{equation}
e_0 = \left[Z_\alpha^{\text{os}}\right]^{1/2} e_{\text{os}}\,,\quad
A_0 = \left[Z_A^{\text{os}}\right]^{1/2} A_{\text{os}}\,.
\label{Coulomb:on-shell}
\end{equation}
By definition, in this scheme the renormalized propagators near the mass shell are free.
In particular,
\begin{equation}
D^{00}(0,\vec{q}) = Z_A^{\text{os}} D^{00}_{\text{os}}(0,\vec{q})\,,\quad
D^{00}_{\text{os}}(0,\vec{q}) \to - \frac{1}{\vec{q}^2}\,,
\label{Coulomb:Don-shell}
\end{equation}
so that
\begin{equation}
Z_\alpha^{\text{os}} = \left[Z_A^{\text{os}}\right]^{-1} = 1 - \Pi(0)\,.
\label{Coulomb:Zon-shell}
\end{equation}

Another widely used renormalization scheme is $\overline{\text{MS}}$:
\begin{equation}
e_0 = Z_\alpha^{1/2}(\alpha(\mu)) e(\mu)\,,\quad
A_0 = Z_A^{1/2}(\alpha(\mu)) A(\mu)\,,
\label{Coulomb:MS}
\end{equation}
where all renormalization constants are minimal
\begin{equation}
Z_i(\alpha) = 1 + \frac{z_1}{\varepsilon} \frac{\alpha}{4\pi}
+ \left(\frac{z_{22}}{\varepsilon^2} + \frac{z_{21}}{\varepsilon}\right) \left(\frac{\alpha}{4\pi}\right)^2
+ \cdots\,,
\label{Coulomb:minimal}
\end{equation}
and
\begin{equation}
\frac{\alpha(\mu)}{4\pi} =
\frac{e^2(\mu)\,\mu^{-2\varepsilon}}{(4\pi)^{d/2}}
e^{-\gamma \varepsilon}
\label{Coulomb:alpha}
\end{equation}
(where $\gamma$ is the Euler constant).
The renormalization constant $Z_A$ is determined by the condition
that the renormalized photon propagator is finite at $\varepsilon \to 0$:
\begin{equation}
D^{00}(0,\vec{q}) = Z_A D^{00}(0,\vec{q};\mu)\,,\quad
D^{00}(0,\vec{q};\mu) = \text{finite}\,.
\label{Coulomb:ZA}
\end{equation}
The potential~(\ref{Coulomb:QED})
\begin{equation*}
U(\vec{q}) = e^2(\mu) D^{00}(0,\vec{q};\mu) Z_\alpha Z_A
\end{equation*}
must be finite, and hence
\begin{equation}
Z_\alpha = Z_A^{-1}\,.
\label{Coulomb:Zalpha}
\end{equation}

The $\overline{\text{MS}}$ $\alpha(\mu)$~(\ref{Coulomb:alpha})
satisfies the renormalization group equation
\begin{equation}
\frac{d\log\alpha(\mu)}{d\log\mu} = - 2 \varepsilon - 2 \beta(\alpha(\mu))\,,\quad
\beta(\alpha(\mu)) = \frac{1}{2} \frac{d\log Z_\alpha(\alpha(\mu))}{d\log\mu}\,,\quad
\beta(\alpha) = \beta_0 \frac{\alpha}{4\pi}
+ \beta_1 \left(\frac{\alpha}{4\pi}\right)^2
+ \cdots
\label{Coulomb:RG}
\end{equation}
Similarly, the $\overline{\text{MS}}$ renormalized photon field~(\ref{Coulomb:MS})
satisfies the renormalization group equation
\begin{equation}
\frac{d A(\mu)}{d\log\mu} = - \frac{1}{2} \gamma_A(\alpha(\mu)) A(\mu)\,,\quad
\gamma_A(\alpha(\mu)) = \frac{d\log Z_A(\alpha(\mu))}{d\log\mu}\,,\quad
\gamma_A(\alpha) = \gamma_{A0} \frac{\alpha}{4\pi}
+ \gamma_{A1} \left(\frac{\alpha}{4\pi}\right)^2
+ \cdots
\label{Coulomb:RGA}
\end{equation}
In QED~(\ref{Coulomb:Zalpha})
\begin{equation}
\beta(\alpha) = - \frac{1}{2} \gamma_A(\alpha)\,.
\label{Coulomb:beta}
\end{equation}

\subsection{Charge decoupling}
\label{S:Dec}

Now we shall discuss the relation between the full theory (QED)
and the low-energy effective theory (QPD) more systematically.
All QPD quantities will be denoted by primes.
In this theory charge is not renormalized:
\begin{equation}
e_0' = e_{\text{os}}' = e'(\mu)\,.
\label{Dec:QPD}
\end{equation}
The macroscopically measured charge is the same in QED and QPD:
\begin{equation}
e_{\text{os}} = e_{\text{os}}'\,.
\label{Dec:Dec}
\end{equation}
The bare charges in the two theories are related by the bare decoupling coefficient:
\begin{equation}
e_0 = \left[\zeta_\alpha^0\right]^{-1/2} e_0'\,,\quad
\zeta_\alpha^0 = \left[Z_\alpha^{\text{os}}\right]^{-1}\,.
\label{Dec:bare}
\end{equation}
The $\overline{\text{MS}}$ charges are related by the renormalized decoupling coefficient:
\begin{equation}
e(\mu) = \left[\zeta_\alpha(\mu)\right]^{-1/2} e'(\mu)\,,\quad
\zeta_\alpha(\mu) = Z_\alpha \zeta_\alpha^0 = \frac{Z_\alpha}{Z_\alpha^{\text{os}}}\,.
\label{Dec:ren}
\end{equation}

The photon self-energy $\Pi(q^2)$ in the 1-loop approximation is
\begin{equation}
\raisebox{-11mm}{\begin{picture}(32,23)
\put(16,11.5){\makebox(0,0){\includegraphics{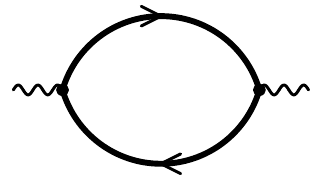}}}
\put(16,0){\makebox(0,0)[b]{$k$}}
\put(16,23){\makebox(0,0)[t]{$k+q$}}
\put(0,10.5){\makebox(0,0)[tl]{$\mu$}}
\put(32,10.5){\makebox(0,0)[tr]{$\nu$}}
\end{picture}} =
i \left( q^2 g_{\mu\nu} - q_\mu q_\nu \right) \Pi(q^2)\,.
\label{Dec:Pi1}
\end{equation}
Expanding in $q$ and using~(\ref{Qedland:V1}), we obtain
\begin{equation}
\Pi(q^2) = - \frac{4}{3} \frac{e_0^2 M_0^{-2\varepsilon}}{(4\pi)^{d/2}}
\Gamma(\varepsilon)
\left( 1 - \frac{d-4}{10} \frac{q^2}{M_0^2} + \cdots \right)\,,
\label{Dec:Pi1res}
\end{equation}
where $M_0$ is the bare electron mass.
The $\mathcal{O}(q^2)$ correction can be used to obtain the contact interaction $c$~(\ref{Coulomb:Uc});
from now on we shall neglect it.
The on-shell charge renormalization constant~(\ref{Coulomb:Zon-shell}) is
\begin{equation}
Z_\alpha^{\text{os}} = \left[\zeta_\alpha^0\right]^{-1} = 1 - \Pi(0)
= 1 + \frac{4}{3} \frac{e_0^2 M_0^{-2\varepsilon}}{(4\pi)^{d/2}} \Gamma(\varepsilon) + \cdots\,,
\label{Dec:Zalos1}
\end{equation}
The renormalized decoupling coefficient $\zeta_\alpha(\mu)$~(\ref{Dec:ren})
must be finite, and hence the $\overline{\text{MS}}$ charge renormalization constant is
\begin{equation}
Z_\alpha(\alpha) = 1 + \frac{4}{3} \frac{\alpha}{4\pi\epsilon} + \cdots
\label{Dec:Zalms1}
\end{equation}
As a free bonus, we have obtained the QED $\beta$ function~(\ref{Coulomb:RG}):
\begin{equation}
\beta_0 = - \frac{4}{3}\,,
\label{Dec:beta0}
\end{equation}
the negative sign corresponds to screening.%
\footnote{The photon self-energy $\Pi(q^2)$ can be written as a dispersion integral with a positive spectral density.
Therefore, the potential~(\ref{Coulomb:QED}) up to 1 loop is a superposition of the Coulomb potential
and Yukawa ones having various radii, with positive weights.
The farther we are from the source, the more Yukawa potentials die out, and the weaker is the interaction.}
The renormalized charge decoupling coefficient expressed via
the $\overline{\text{MS}}$ renormalized $\alpha(\mu)$ and electron mass $M(\mu)$ is
\begin{equation}
\left[\zeta_\alpha(\mu)\right]^{-1} = 1
+ \frac{4}{3} \left[ \left(\frac{\mu}{M(\mu)}\right)^{2\varepsilon} e^{\gamma\epsilon} \Gamma(1+\varepsilon) - 1 \right]
\frac{\alpha(\mu)}{4\pi\varepsilon} + \cdots
\to 1 + \frac{4}{3} \frac{\alpha(\mu)}{4\pi} L + \cdots\,,
\label{Dec:z1}
\end{equation}
where
\begin{equation}
L = 2 \log\frac{\mu}{M(\mu)}\,.
\label{Dec:L}
\end{equation}

\begin{figure}[ht]
\begin{center}
\begin{picture}(106,17)
\put(16,8.5){\makebox(0,0){\includegraphics{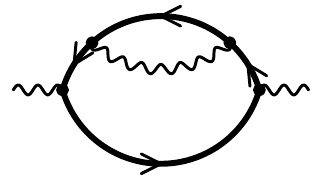}}}
\put(53,8.5){\makebox(0,0){\includegraphics{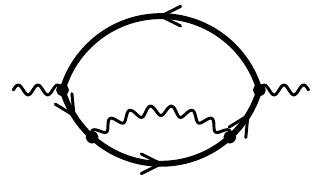}}}
\put(90,8.5){\makebox(0,0){\includegraphics{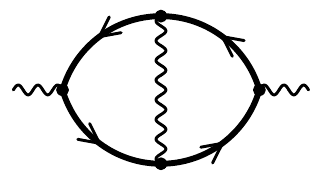}}}
\end{picture}
\end{center}
\caption{The 2-loop photon self-energy.}
\label{F:Pi2}
\end{figure}

The photon self-energy $\Pi(0)$ at 2 loops (Fig.~\ref{F:Pi2})
can be calculated using~(\ref{Qedland:V2}):
\begin{align}
&Z_\alpha^{\text{os}} = \left[\zeta_\alpha^0\right]^{-1} = 1 - \Pi(0) = 1
+ \frac{4}{3} \frac{e_0^2 M_0^{-2\varepsilon}}{(4\pi)^{d/2}} \Gamma(\varepsilon)
+ \frac{2}{3} \frac{(d-4)(5d^2-33d+34)}{d(d-5)}
\left(\frac{e_0^2 M_0^{-2\varepsilon}}{(4\pi)^{d/2}}
\Gamma(\varepsilon)\right)^2
+ \cdots
\nonumber\\
&{} = 1 + \frac{4}{3} \frac{\alpha(\mu)}{4\pi\varepsilon} e^{L\varepsilon}
\left(1 + \frac{\pi^2}{12} \varepsilon^2 + \cdots\right)
Z_\alpha(\alpha(\mu)) Z_m^{-2\varepsilon}(\alpha(\mu))
- \varepsilon \left(6 - \frac{13}{3} \varepsilon + \cdots\right)
\left(\frac{\alpha(\mu)}{4\pi\varepsilon}\right)^2 e^{2L\varepsilon}
+ \cdots
\label{Dec:Zalos2}
\end{align}

In the 1-loop term we need the 1-loop $Z_\alpha$~(\ref{Dec:Zalms1})
and the 1-loop $\overline{\text{MS}}$ mass renormalization constant $Z_m$
defined by
\begin{equation}
M_0 = Z_m(\alpha(\mu)) M(\mu) = Z_m^{\text{os}} M_{\text{os}}\,.
\label{Dec:Zm}
\end{equation}
The 1-loop on-shell renormalization constant $Z_m^{\text{os}}$
can be easily calculated using the on-shell integrals
\begin{equation}
\begin{split}
&\raisebox{-8.5mm}{\begin{picture}(42,18)
\put(21,9){\makebox(0,0){\includegraphics{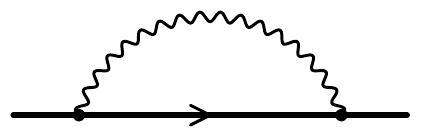}}}
\put(21,18){\makebox(0,0)[t]{$k$}}
\put(21,0){\makebox(0,0)[b]{$k+P$}}
\put(21,10){\makebox(0,0)[b]{$n_2$}}
\put(21,7){\makebox(0,0)[t]{$n_1$}}
\end{picture}} =
\frac{1}{i \pi^{d/2}} \int \frac{d^d k}{D_1^{n_1} D_2^{n_2}} = M^{d-2(n_1+n_2)} M(n_1,n_2)\,,\quad
P^2 = M^2\,,\\
&D_1 = M^2 - (k+P)^2\,,\quad
D_2 = - k^2\,,\quad
M(n_1,n_2) = \frac{\Gamma(d - n_1 - 2 n_2) \Gamma\left(n_1 + n_2 - \frac{d}{2}\right)}%
{\Gamma(n_1) \Gamma(d - n_1 - n_2)}\,.
\end{split}
\label{Dec:M1}
\end{equation}
The result is
\begin{equation}
Z_m^{\text{os}} = 1
- \frac{d-1}{d-3} \frac{e_0^2 M_0^{-2\varepsilon}}{(4\pi)^{d/2}} \Gamma(\varepsilon)
+ \cdots
\label{Dec:Zmos1}
\end{equation}
Both $M_{\text{os}}$ and $M(\mu)$ are finite at $\varepsilon\to0$, hence
\begin{equation}
Z_m(\alpha) = 1 - 3 \frac{\alpha}{4\pi\varepsilon} + \cdots
\label{Dec:Zmms1}
\end{equation}

Now we have $Z_\alpha^{\text{os}} = \left[\zeta_\alpha^0\right]^{-1}$
expressed via the renormalized quantities $\alpha(\mu)$, $M(\mu)$.
The renormalized decoupling coefficient~(\ref{Dec:ren}) must be finite,
and hence we obtain the 2-loop $\overline{\text{MS}}$ charge renormalization constant
\begin{equation}
Z_\alpha^{-1}(\alpha) = 1 - \frac{4}{3} \frac{\alpha}{4\pi\varepsilon}
- 2 \varepsilon \left(\frac{\alpha}{4\pi\varepsilon}\right)^2
+ \cdots
\label{Dec:Zalms2}
\end{equation}
(by the way, it gives $\beta_1 = -4$).
The renormalized charge decoupling coefficient is
\begin{equation}
\zeta_\alpha^{-1}(\mu) = 1
+ \frac{4}{3} \left[L + \left(\frac{L^2}{2} + \frac{\pi^2}{12}\right) \varepsilon\right] \frac{\alpha(\mu)}{4\pi}
+ \left(- 4 L + \frac{13}{3}\right) \left(\frac{\alpha(\mu)}{4\pi}\right)^2
+ \cdots
\label{Dec:z2}
\end{equation}
If we define $\bar{M}$ as the root of the equation
\begin{equation}
M(\bar{M}) = \bar{M}\,,
\label{Dec:barM}
\end{equation}
then $L=0$ at $\mu = \bar{M}$, and
\begin{equation}
\zeta_\alpha^{-1}(\bar{M})
= 1 + \frac{\pi^2}{9} \varepsilon \frac{\alpha(\bar{M})}{4\pi}
+ \frac{13}{3} \left(\frac{\alpha(\bar{M})}{4\pi}\right)^2
+ \cdots
\label{Dec:zMb}
\end{equation}
Note that the 1-loop $\mathcal{O}(\varepsilon)$ term is needed in 2-loop calculations,
because it can get multiplied by an $1/\varepsilon$ divergent 1-loop integral.

Alternatively, we can use use $\mu = M_{\text{os}}$.
From~(\ref{Dec:Zmos1}) and~(\ref{Dec:Zmms1}) we have
\begin{equation*}
\frac{M(\mu)}{M_{\text{os}}} = 1
- 6 \left( \log\frac{\mu}{M_{\text{os}}} + \frac{2}{3} \right)
\frac{\alpha}{4\pi} + \cdots\,,
\end{equation*}
so that
\begin{equation*}
L = 8 \frac{\alpha}{4\pi}\,,
\end{equation*}
and we obtain
\begin{equation}
\zeta_\alpha^{-1}(M_{\text{os}})
= 1 + \frac{\pi^2}{9} \varepsilon \frac{\alpha(M_{\text{os}})}{4\pi}
+ 15 \left(\frac{\alpha(M_{\text{os}})}{4\pi}\right)^2
+ \cdots
\label{Dec:zMos}
\end{equation}
For any $\mu=M(1+\mathcal{O}(\alpha))$,
$\zeta_\alpha(\mu) = 1 + \mathcal{O}(\varepsilon) \alpha + \mathcal{O}(\alpha^2)$;
if we choose, say, $\mu = 2 M$ or $\mu = M/2$,
there will be a finite correction of order $\alpha$.

More details about decoupling can be found, e.\,g., in the review~\cite{G:13}.

\subsection{Qedland again}
\label{S:NP}

Physicists in Qedland suspect that QED is also only a low-energy effective theory.
We know that they are right, and muons exist.%
\footnote{In our real world $M_\pi\sim M_\mu$;
for simplicity we shall assume that pions and other light hadrons don't exist.}
There are two ways in which they can search for new physics:
\begin{itemize}
\item by increasing the energy of their $e^+ e^-$ colliders
in the hope to produce pairs of new particles;
\item by performing high-precision experiments at low energies
(e.\,g., by measuring the electron magnetic moment).
\end{itemize}
New physics can produce new local interactions of photons,
electrons, and positrons at low energies,
which should be included in the effective QED Lagrangian.
We were lucky that the scale of new physics in QED, the muon mass $M$,
is far away from the electron mass: $M \gg m_e$.
Contributions of muon loops to low-energy processes are also strongly suppressed
by powers of $\alpha$.
Therefore, the prediction for the electron magnetic moment
from the pure QED Lagrangian (without nonrenormalizable corrections)
is in good agreement with experiment.

After this spectacular success of the simplest Dirac equation
(without the Pauli term) for electrons,
physicists expected that the same holds for the proton,
and its magnetic moment is $e/(2 M_p)$.
No luck here.
This shows that the picture of the proton
as a point-like structureless particle
is a poor approximation already at the energy scale $M_p$.

\subsection{Heavy flavors in QCD}
\label{S:QCD}

In QED, effects of decoupling of muon loops are tiny.
Also, pion pairs become important at about the same energies
as muon pairs, so that QED with electrons and muons
is a model with a narrow region of applicability.
In QCD, decoupling of heavy flavors is fundamental and omnipresent.
It would be a huge mistake to use the full 6-flavor QCD
at characteristic energies of order GeV:
running of $\alpha_s(\mu)$ and other quantities would be grossly inadequate,
convergence of perturbative series would be awful because of large logarithms.
In most cases, everybody working in QCD uses an effective low-energy theory,
where a few heaviest flavors have been eliminated
(even if he does not know that he speaks prose).

Let's consider QCD with a single heavy flavor having mass $M$;
for simplicity, all other flavors are supposed to be massless.
Then the behavior of light quarks and gluons at low momenta $p_i\ll M$
is described by the low-energy effective theory.
Its Lagrangian is the usual QCD Lagrangian
(of course, without the heavy-quark field)
plus higher-dimensional terms
(whose coefficients are suppressed by powers of $1/M$).
Power corrections to the Lagrangian first appear at dimension 6.

The full-QCD coupling $\alpha_s^{(n_l+1)}(\mu)$ is related
to the effective-theory coupling $\alpha_s^{(n_l)}(\mu)$
by the decoupling coefficient
\begin{equation}
\alpha_s^{(n_l+1)}(\mu) = \zeta_\alpha^{-1}(\mu) \alpha_s^{(n_l)}(\mu)\,.
\label{DQCD:alpha}
\end{equation}
At $\mu = \bar{M}$ it is given by~\cite{BW:82,LRV:95,CKS:98}
\begin{equation}
\zeta_\alpha^{-1}(\bar{M})
= 1 + \left( \frac{13}{3} C_F - \frac{32}{9} C_A \right) T_F
\left(\frac{\alpha_s(\bar{M})}{4\pi}\right)^2 + \cdots
\label{DQCD:zeta}
\end{equation}
Here the $C_F$ term can be obtained from the QED result~(\ref{Dec:zMb})
by inserting the obvious color factors;
the $C_A$ term needs a new calculation.
The decoupling coefficient $\zeta_\alpha(\mu)$ for any $\mu$
can be found by solving the renormalization group equation
\begin{equation}
\frac{d\log\zeta_\alpha(\mu)}{d\log\mu}
- 2 \beta^{(n_l+1)}(\alpha_s^{(n_l+1)}(\mu))
+ 2 \beta^{(n_l)}(\alpha_s^{(n_l)}(\mu)) = 0
\label{DQCD:RG}
\end{equation}
with the initial condition~(\ref{DQCD:zeta}).

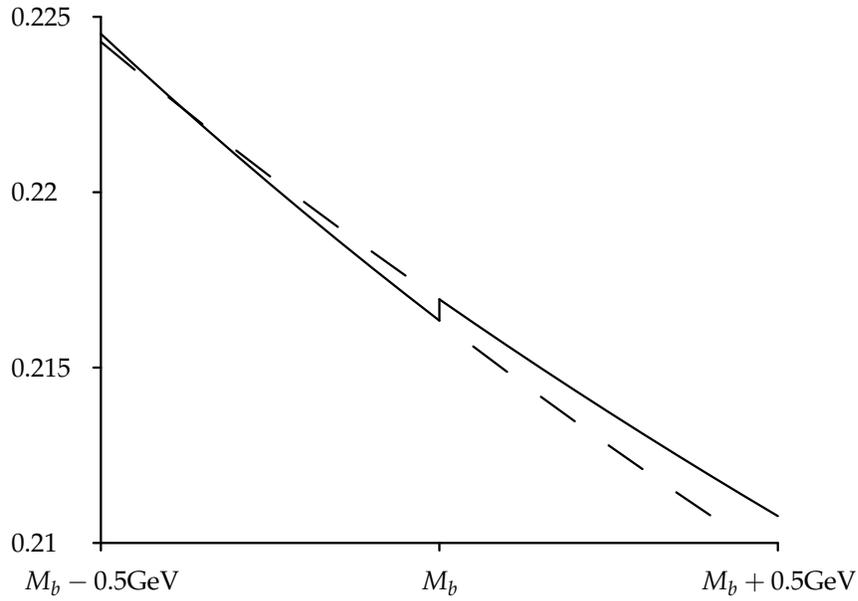
\begin{figure}[ht]
\begin{center}
\begin{picture}(100,80)
\put(50,40){\makebox(0,0){\includegraphics{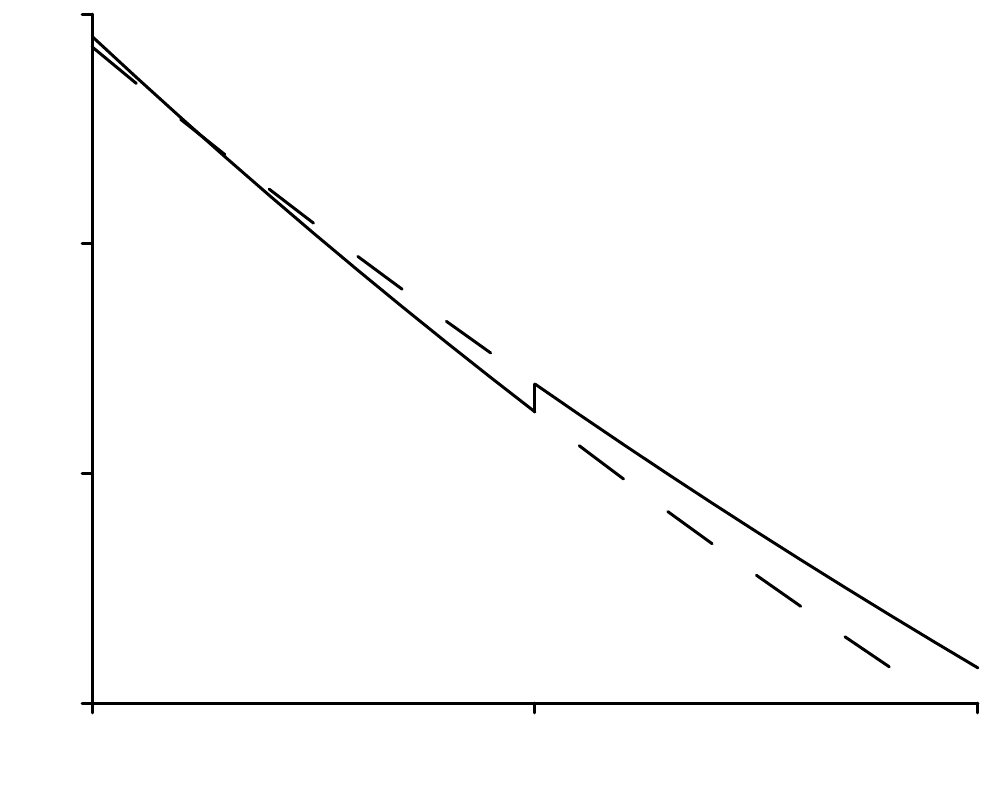}}}
\put(54,2){\makebox(0,0)[b]{$M_b$}}
\put(9,2){\makebox(0,0)[b]{$M_b-0.5\text{GeV}$}}
\put(99,2){\makebox(0,0)[b]{$M_b+0.5\text{GeV}$}}
\put(-3,9){\makebox(0,0)[l]{0.21}}
\put(-3,32.333333){\makebox(0,0)[l]{0.215}}
\put(-3,55.666667){\makebox(0,0)[l]{0.22}}
\put(-3,79){\makebox(0,0)[l]{0.225}}
\end{picture}
\end{center}
\caption{$\alpha_s^{(5)}(\mu)$ and $\alpha_s^{(4)}(\mu)$}
\label{F:as}
\end{figure}

The QCD running coupling $\alpha_s(\mu)$ not only runs when $\mu$ varies;
it also jumps when crossing heavy-flavor thresholds.
The behavior of $\alpha_s(\mu)$ near $M_b$ is shown in Fig.~\ref{F:as}.
At $\mu>m_b$, the correct theory is the full 5-flavor QCD
($\alpha_s^{(5)}(\mu)$, the solid line);
at $\mu<m_b$, the correct theory is the effective low-energy 4-flavor QCD
($\alpha_s^{(4)}(\mu)$, the solid line);
the jump at $\mu=M_b$ is shown.
Of course, both curves can be continued across $M_b$ (dashed lines),
and it is inessential at which particular $\mu\sim M_b$
we switch from one theory to the other one.
However, the on-shell mass $M_b^{\text{os}}$
(or any other mass which differs from it by $\mathcal{O}(\alpha_s)$,
such as, e.g., $\bar{M}_b$)
is most convenient, because the jump is small, $\mathcal{O}(\alpha_s^3)$.
For, say, $\mu=2M_b$ or $\mu=M_b/2$, it would be $\mathcal{O}(\alpha_s^2)$.

\subsection{Method of regions}
\label{S:Reg}

This method (see the textbook~\cite{S:02}) provides an alternative insight
to effective field theories.
It allows one to calculate diagrams expanded in small ratios of scales directly.
It is convenient when you need to calculate a small number of diagrams
expanded up to a high order in a small parameter,
because effective Lagrangians quickly become very long at such high orders.
On the other hand, effective Lagrangians are applicable for all processes,
and are useful for investigation general properties (symmetries, factorization).

Let's consider the vacuum integral
\begin{equation*}
I =
\raisebox{-13.5mm}{\begin{picture}(22,28)
\put(11,14){\makebox(0,0){\includegraphics{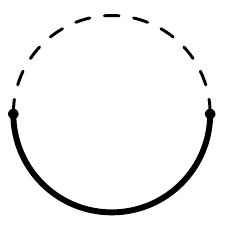}}}
\put(11,0){\makebox(0,0)[b]{$M$}}
\put(11,28){\makebox(0,0)[t]{$m$}}
\end{picture}} =
\int \frac{d^d k}{i\pi^{d/2}} \frac{1}{(M^2-k^2-i0) (m^2-k^2-i0)}
\end{equation*}
with two masses $M$ and $m$ at $M\gg m$ and $d=2$.
It contains neither ultraviolet (UV) nor infrared (IR) divergences.
After Wick rotation to Euclidean momentum space it becomes
\begin{equation}
I = \int \frac{d^d k_E}{\pi^{d/2}} \frac{1}{(k_E^2+M^2) (k_E^2+m^2)}\,.
\label{Reg:Idef}
\end{equation}
Of course, in this simple example it is easy to obtain the exact solution.
We use partial fraction decomposition
\begin{align*}
&\raisebox{-10.25mm}{\includegraphics{reg1.pdf}}
= \frac{1}{M^2-m^2} \Biggl[
- \raisebox{-10.25mm}{\includegraphics{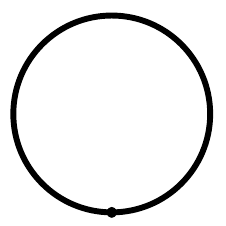}}
+ \raisebox{-10.25mm}{\includegraphics{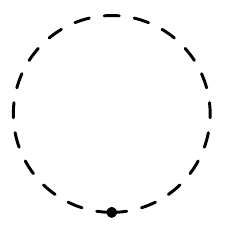}}
\Biggr]\,,\\
&I = \frac{1}{M^2-m^2} \int \frac{d^dk_E}{\pi^{d/2}}
\left[ - \frac{1}{k_E^2+M^2} + \frac{1}{k_E^2+m^2} \right]\,.
\end{align*}
These two integrals, taken separately, diverge;
therefore, we use dimensional regularization ($d=2-2\varepsilon$) and obtain
\begin{equation*}
I = - \Gamma(\varepsilon) \frac{M^{-2\varepsilon} - m^{-2\varepsilon}}{M^2 - m^2}\,.
\end{equation*}
This result is finite at $\varepsilon\to0$:
\begin{equation}
I = \frac{\log(M^2/m^2)}{M^2 - m^2}
= \frac{1}{M^2} \log\frac{M^2}{m^2}
\left[ 1 + \frac{m^2}{M^2} + \frac{m^4}{M^4} + \cdots \right]\,.
\label{Reg:exact}
\end{equation}

It is easier to obtain this result using the prescription known as the method of regions.
The integral~(\ref{Reg:Idef}) is written at the sum of contributions of two regions,
the hard one and the soft one:
\begin{equation*}
I = I_h + I_s\,.
\end{equation*}
In the hard region $k_E\sim M$;
in the soft one $k_E\sim m$.
The integrand is expanded to Taylor series
in accordance to these power counting rules in each region.
After that, the integral is taken over the full ($d$-dimensional) space.

In the hard region ($k_E\sim M$) we have
\begin{equation*}
I_h = \int \frac{d^dk_E}{\pi^{d/2}}
T_h \frac{1}{(k_E^2 + M^2) (k_E^2 + m^2)}\,,
\end{equation*}
where the operator $T_h$ expands the integrand in small parameter(s)
counting $k_E$ as a quantity of order $M$:
\begin{equation*}
T_h \frac{1}{(k_E^2 + M^2) (k_E^2 + m^2)}
= \frac{1}{k_E^2 + M^2} \frac{1}{k_E^2}
\left[ 1 - \frac{m^2}{k_E^2} + \frac{m^4}{k_E^4} - \cdots \right]\,.
\end{equation*}
Calculating the loop integrals, we arrive at
\begin{equation}
I_h = - \frac{M^{-2\varepsilon}}{M^2} \Gamma(\varepsilon)
\left[ 1 + \frac{m^2}{M^2} + \frac{m^4}{M^4} + \cdots \right]\,.
\label{Reg:hard}
\end{equation}
The result is a Taylor series in $m$.
Each loop integral is IR divergent;
it contains a single scale $M$,
and hence, by dimensions counting, is proportional to $M^{-2\varepsilon}$.

In the soft region ($k_E\sim m$) we have
\begin{equation*}
I_s = \int \frac{d^dk_E}{\pi^{d/2}}
T_s \frac{1}{(k_E^2 + M^2) (k_E^2 + m^2)}\,,
\end{equation*}
where $T_s$ counts $k_E$ as a quantity of order $m$:
\begin{equation*}
T_s \frac{1}{(k_E^2 + M^2) (k_E^2 + m^2)}
= \frac{1}{M^2} \frac{1}{k_E^2 + m^2}
\left[ 1 - \frac{k_E^2}{M^2} + \frac{k_E^4}{M^4} - \cdots \right]\,,
\end{equation*}
and we obtain
\begin{equation}
I_s = \frac{m^{-2\varepsilon}}{M^2} \Gamma(\varepsilon)
\left[ 1 + \frac{m^2}{M^2} + \frac{m^4}{M^4} + \cdots \right]\,.
\label{Reg:soft}
\end{equation}
The result is a Taylor series in $1/M$.
Each loop integral is UV divergent;
it contains a single scale $m$,
and hence, by dimensions counting, is proportional to $m^{-2\varepsilon}$.

The sum of the hard contribution~(\ref{Reg:hard}) and the soft one~(\ref{Reg:soft})
produces the complete result~(\ref{Reg:exact}).
IR divergences in the hard region cancel UV divergences in the soft one.

In this simple case it is easy to prove this prescription~\cite{J:11}.
Let's introduce some boundary $\Lambda$ such that $m\ll\Lambda\ll M$,
and write
\begin{equation*}
I = \int_{k_E>\Lambda} \frac{d^dk_E}{\pi^{d/2}}
\frac{1}{(k_E^2 + M^2) (k_E^2 + m^2)}
+ \int_{k_E<\Lambda} \frac{d^dk_E}{\pi^{d/2}}
\frac{1}{(k_E^2 + M^2) (k_E^2 + m^2)}\,.
\end{equation*}
We may apply $T_h$ to the first integrand and $T_s$ to the second one:
\begin{equation*}
I = \int_{k_E>\Lambda} \frac{d^dk_E}{\pi^{d/2}}
T_h \frac{1}{(k_E^2 + M^2) (k_E^2 + m^2)}
+ \int_{k_E<\Lambda} \frac{d^dk_E}{\pi^{d/2}}
T_s \frac{1}{(k_E^2 + M^2) (k_E^2 + m^2)}\,.
\end{equation*}
For each of these two integrals, we add and subtract the integral
over the remaining part of the $k_E$ space:
\begin{equation*}
I = I_h + I_s - \Delta I\,,
\end{equation*}
where 
\begin{equation*}
\Delta I = \int_{k_E<\Lambda} \frac{d^dk_E}{\pi^{d/2}}
T_h \frac{1}{(k_E^2 + M^2) (k_E^2 + m^2)}
+ \int_{k_E>\Lambda} \frac{d^dk_E}{\pi^{d/2}}
T_s \frac{1}{(k_E^2 + M^2) (k_E^2 + m^2)}
\end{equation*}
are the integrals of the two expansions over the ``wrong'' parts of the $k_E$ space.
In these wrong regions, we may apply the other expansion in addition to the existing one:
\begin{equation*}
\Delta I = \int_{k_E<\Lambda} \frac{d^dk_E}{\pi^{d/2}}
T_s T_h \frac{1}{(k_E^2 + M^2) (k_E^2 + m^2)}
+ \int_{k_E>\Lambda} \frac{d^dk_E}{\pi^{d/2}}
T_h T_s \frac{1}{(k_E^2 + M^2) (k_E^2 + m^2)}\,.
\end{equation*}
But this is the integral over the full $k_E$ space:
\begin{equation*}
\Delta I = \int \frac{d^dk_E}{\pi^{d/2}}
T_s T_h \frac{1}{(k_E^2 + M^2) (k_E^2 + m^2)}\,,
\end{equation*}
because the Taylor expansion operators commute:
\begin{align*}
&T_s T_h \frac{1}{(k_E^2 + M^2) (k_E^2 + m^2)}
= T_h T_s \frac{1}{(k_E^2 + M^2) (k_E^2 + m^2)}\\
&{} = \frac{1}{M^2 k_E^2}
\left[ 1 - \frac{m^2}{k_E^2} + \frac{m^4}{k_E^4} - \cdots \right]
\left[ 1 - \frac{k_E^2}{M^2} + \frac{k_E^4}{M^4} - \cdots \right]\,.
\end{align*}
And hence
\begin{equation*}
\Delta I = 0\,,
\end{equation*}
because each integral has no scale.

Let's stress that the method of regions works thanks to dimensional regularization.
Taylor expansions in each region (i.e., for each set of power counting rules)
must be performed completely, up to the end;
otherwise, integrals in the defect $\Delta I$ can contain some scale(s),
and hence not vanish.

Let's consider QCD with $n_l$ light flavors and a heavy flavor of mass $M$,
and calculate some scattering amplitude of light quarks and gluons
with low characteristic energies $\sim E \ll M$.
We can use the method of regions.
For each diagram, some subdiagrams will be hard
(characteristic momenta $\sim M$;
in particular, all heavy-quark loops are in such subdiagrams);
light-quark and gluon lines connecting these hard subdiagrams will be soft
(characteristic momenta $\sim E$).
In the EFT language, these hard subdiagrams are local interactions
present in the effective Lagrangian;
the overall soft diagram is calculated according to the Feynman rules
of the low-energy theory.

\section{Bloch--Nordsieck EFT, HQET}
\label{S:BN}

In the effective field theories we have considered previously
heavy particles are completely absent in the effective Lagrangians.
There is another kind of effective theories
in which heavy particles are retained,
but move non-relativistically in some reference frame
and can be described in a simplified way.
This effective theory has been first constructed by Bloch and Nordsieck~\cite{BN:37}
to describe interaction of an electron with soft photons.
Its non-abelian version is called heavy quark effective theory (HQET),
see, e.\,g., \cite{N:94,MW:00,G:04}.
It is used to describe properties of hadrons with a single heavy quark in QCD.

\subsection{Heavy electron effective theory}
\label{S:HEET}

Photonia has imported a single electron from Qedland,
and physicists are studying its interaction with soft photons
(both real and virtual) which they can produce and detect so well.
The aim is to construct a theory describing states
with a single electron plus soft photon fields.

The ground state (``vacuum'') of the theory is the electron at rest
(and no photons).
It is natural to define its energy to be $0$.
When the electron has momentum $\vec{p}$, its energy is
\begin{equation}
\varepsilon(\vec{p}\,) = \frac{\vec{p}\,^2}{2M}\,,
\label{HEET:kinetic}
\end{equation}
where $M$ is the electron mass (in the on-shell renormalization scheme),
our large mass scale.
The electron velocity is
\begin{equation}
\vec{v} = \frac{\partial\varepsilon(\vec{p}\,)}{\partial\vec{p}}
= \frac{\vec{p}}{M}\,.
\label{HEET:velocity}
\end{equation}

At the leading ($0$-th) order in $1/M$,
the mass shell of the free electron is
\begin{equation}
\varepsilon(\vec{p}\,) = 0\,.
\label{HEET:mshell}
\end{equation}
At this order, the electron velocity is
\begin{equation}
\vec{v} = \frac{\partial\varepsilon(\vec{p}\,)}{\partial\vec{p}}
= \vec{0}\,.
\label{HEET:zero}
\end{equation}
The electron does not move; it always stays in the point
where it has been put initially.
The Lagrangian
\begin{equation}
L = h^+ i \partial_0 h\,,
\label{HEET:L0}
\end{equation}
where $h$ is the 2-component spinor electron field,
leads to the equation of motion
\begin{equation}
i \partial_0 h = 0\,.
\label{HEET:EOM0}
\end{equation}
This means that the energy of an on-shell electron
is $\varepsilon=0$.
Thus the Lagrangian~(\ref{HEET:L0})
reproduces the mass shell~(\ref{HEET:mshell}),
and can be used to describe the free electron
at the leading order in $1/M$.

The electron has charge $-e$.
Therefore, when placed in an external electromagnetic field,
it has energy
\begin{equation}
\varepsilon = - e A_0
\label{HEET:E}
\end{equation}
instead of~(\ref{HEET:mshell}).
Therefore, the equation of motion is
\begin{equation}
i D_0 h = 0
\label{HEET:EOM}
\end{equation}
instead of~(\ref{HEET:EOM0}), where
\begin{equation}
D_\mu = \partial_\mu - i e A_\mu
\label{HEET:covariant}
\end{equation}
is the covariant derivative.
It can be obtained from the HEET Lagrangian~\cite{EH:90}
\begin{equation}
L = h^+ i D_0 h\,.
\label{HEET:L}
\end{equation}
This Lagrangian is not Lorentz-invariant.
It is invariant with respect to the gauge transformation
\begin{equation}
A_\mu \to A_\mu + \partial_\mu \alpha(x)\,,\quad
h \to e^{i e \alpha(x)} h\,.
\label{HEET:gauge}
\end{equation}

Of course, the full Lagrangian is the sum of~(\ref{HEET:L})
and the Lagrangian of the photon field.
This gives the equation of motion for the electromagnetic field
\begin{equation}
\partial_\mu F^{\mu\nu} = j^\nu\,,
\label{HEET:Maxwell}
\end{equation}
where the current $j^\mu$ has only $0$-th component
\begin{equation}
j^0 = - e h^+ h
\label{HEET:j}
\end{equation}
(the interaction term in the Lagrangian~(\ref{HEET:L}) is $-j^\mu A_\mu$).
The electron produces the Coulomb field.

At the leading order in $1/M$,
the electron spin does not interact with electromagnetic field.
We can rotate it without affecting physics.
Speaking more formally, the Lagrangian~(\ref{HEET:L}) has,
in addition to the $U(1)$ symmetry $h\to e^{i\alpha}h$,
also the $SU(2)$ spin symmetry~\cite{IW:90}:
it is invariant with respect to transformations
\begin{equation}
h \to U h\,,
\label{HEET:SU2}
\end{equation}
where $U$ is a $SU(2)$ matrix ($U^+ U=1$).

In fact, the electron has magnetic moment $\vec{\mu}=\mu\vec{\sigma}$
proportional to its spin $\vec{s}=\vec{\sigma}/2$,
and this magnetic moment interacts with magnetic field:
the interaction Hamiltonian is $-\vec{\mu}\cdot\vec{B}$.
But by dimensionality the magnetic moment $\mu\sim e/M$,
and this interaction only appears at the level of $1/M$ corrections.
Namely, $\mu=-\mu_B$ (up to small radiative corrections), where
\begin{equation}
\mu_B = \frac{e}{2M}
\label{HEET:mu}
\end{equation}
is the Bohr magneton.
The Lagrangian thus has an additional term
describing this magnetic interaction,
\begin{equation}
L_m = - \frac{e}{2M} h^+ \vec{B}\cdot\vec{\sigma} h\,.
\label{HEET:Lm}
\end{equation}
This term violates the $SU(2)$ spin symmetry at the $1/M$ level.

If we assume that there are $n_f$ flavors of heavy fermions,
\begin{equation}
L = \sum_{i=1}^{n_f} h_i^+ i D_0 h_i\,,
\label{HEET:nf}
\end{equation}
then the Lagrangian has $U(1)\times SU(2n_f)$ symmetry
(even when the masses $M_i$ are different).
The spin-flavor symmetry is broken at the $1/M_i$ level
by both the kinetic-energy term and the magnetic-interaction term.

At the leading order in $1/M$,
not only the spin direction but also its magnitude is irrelevant.
We can, for example, switch the electron spin off:
\begin{equation}
L = \varphi^* i D_0 \varphi\,,
\label{HEET:spin0}
\end{equation}
where $\varphi$ is a scalar field (with charge $-e$).
This is the most convenient form of the Lagrangian in all cases
when we are not interested in $1/M$ corrections.
If we consider the scalar and the spinor fields together,
\begin{equation}
L = \varphi^* i D_0 \varphi + h^+ i D_0 h\,,
\label{HEET:spins}
\end{equation}
then this Lagrangian has $U(1)\times SU(3)$ symmetry~\cite{GW:90}.
The superflavor $SU(3)$ symmetry contains, in addition to $SU(2)$
spin transformations~(\ref{HEET:SU2})
and phase rotations $\varphi\to e^{2i\alpha}\varphi$, $h\to e^{-i\alpha}h$,
also transformations which mix spin-0 and spin-$\frac{1}{2}$ fields.
In the infinitesimal form,
\begin{equation}
\delta \left(\begin{array}{c}\varphi\\h\end{array}\right) =
i \left(\begin{array}{cc}0&\varepsilon^+\\\varepsilon&0\end{array}\right)
\left(\begin{array}{c}\varphi\\h\end{array}\right)\,,
\label{HEET:super}
\end{equation}
where $\varepsilon$ is an infinitesimal spinor parameter.
So, this $SU(3)$ is a supersymmetry group.
If we want, we can consider, e.\,g., spins $\frac{1}{2}$ and 1;
the corresponding Lagrangian has $SU(5)$ superflavor symmetry.
The superflavor symmetry is broken at the $1/M$ level
by the magnetic-interaction term in the Lagrangian~(\ref{HEET:Lm}).

\subsection{Feynman rules}
\label{S:Feyn}

For now, we are working at the leading order in $1/M$.
The HEET Lagrangian expressed via the bare fields and parameters is
\begin{equation}
L = \varphi_0^* i D_0 \varphi_0
- \frac{1}{4} F_{0\mu\nu} F_0^{\mu\nu}
- \frac{1}{2 a_0} (\partial_\mu A_0^\mu)^2\,,\quad
D_\mu = \partial_\mu - i e_0 A_{0\mu}\,.
\label{Feyn:L}
\end{equation}
It gives the usual photon propagator.
From the free electron part $\varphi_0^* i \partial_0 \varphi_0$
we obtain the momentum-space free electron propagator
\begin{equation}
\raisebox{-3.5mm}{\begin{picture}(22,6.5)
\put(11,4.5){\makebox(0,0){\includegraphics{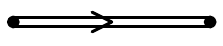}}}
\put(11,0){\makebox(0,0)[b]{$p$}}
\end{picture}}
= i S_0(p)\,,\qquad
S_0(p) = \frac{1}{p_0+i0}\,.
\label{Feyn:Sp}
\end{equation}
It depends only on $p_0$, not on $\vec{p}$.
If we use the spin-$\frac{1}{2}$ field $h_0$
instead of the spin-0 field $\varphi_0$,
then the unit $2\times2$ spin matrix is assumed here.
The coordinate-space propagator is its Fourier transform:
\begin{equation}
\raisebox{-3.5mm}{\begin{picture}(22,6.5)
\put(11,4.5){\makebox(0,0){\includegraphics{fr2.pdf}}}
\put(1,0){\makebox(0,0)[b]{$0$}}
\put(21,0){\makebox(0,0)[b]{$x$}}
\end{picture}}
= i S_0(x)\,,\qquad
S_0(x) = S_0(x_0) \delta(\vec{x}\,)\,,\qquad
S_0(t) = - i \theta(t)\,.
\label{Feyn:Sx}
\end{equation}
The infinitely heavy (static) electron does not move:
it always stays at the point where it has been placed initially.
Alternatively, instead of Fourier-transforming~(\ref{Feyn:Sp}),
we can obtain~(\ref{Feyn:Sx}) by direct solving the equation
\begin{equation}
i \partial_0 S_0(x) = \delta(x)
\label{Feyn:propeq0}
\end{equation}
for the free $x$-space propagator.
Finally, the interaction term $e_0 \varphi_0^* \varphi_0 A_0^0$
in~(\ref{Feyn:L}) produces the vertex
\begin{equation}
\raisebox{-1mm}{\begin{picture}(22,15)
\put(11,6.5){\makebox(0,0){\includegraphics{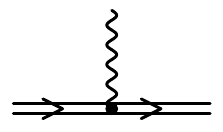}}}
\put(11,15){\makebox(0,0)[t]{$\mu$}}
\end{picture}}
= i e_0 v^\mu\,,
\label{Feyn:vert}
\end{equation}
where
\begin{equation}
v^\mu = (1,\vec{0}\,)
\label{Feyn:v}
\end{equation}
is the 4-velocity of our laboratory frame
(in which the electron is nearly at rest all the time).

The static field $\varphi_0$ (or $h_0$)
describes only particles, there are no antiparticles.
Therefore, there are no pair creation and annihilation
(even virtual).
In other words, there are no loops formed by propagators
of the static electron.
The electron propagates only forward in time~(\ref{Feyn:Sx});
the product of $\theta$ functions along a loop vanishes.
We can also see this in momentum space:
all poles of the propagators~(\ref{Feyn:Sp}) in such a loop
are in the lower $p_0$ half-plane,
and closing the integration contour upwards, we get 0.

It is easy to find the propagator of the static electron
an an arbitrary external electromagnetic field $A^\mu(x)$.
It satisfies the equation
\begin{equation}
i D_0 S(x,x') = (i \partial_0 + e_0 A^0(x)) S(x,x') = \delta(x-x')
\label{Feyn:propeq}
\end{equation}
instead of~(\ref{Feyn:propeq0})
(the derivative $\partial_0$ acts on $x$).
Its solution is
\begin{equation}
S(x,x') = S(x_0,x_0') \delta(\vec{x}-\vec{x}\,')\,,\qquad
S(x_0,x_0') = S_0(x_0-x_0') W(x_0,x_0')\,,
\label{Feyn:SA}
\end{equation}
where
\begin{equation}
W(x_0,x_0') = \exp i e_0 \int\limits_{x_0'}^{x_0} A^\mu(t,\vec{x}\,) v_\mu dt
\label{Feyn:W}
\end{equation}
is the straight Wilson line from $x'$ to $x$ (along $v$).
The same formula can be used when the electromagnetic field
is quantum (operator $A_0^\mu(x)$),
but the exponent~(\ref{Feyn:W}) has to be path-ordered:
operators referring to earlier points (along the path)
are placed to the right from those for later points.
This is usually denoted by $P\exp$;
when the path is directed to the future,
$P$-ordering coincides with $T$-ordering.
The Wilson line has a useful property
\begin{equation}
D_0 W(x,x') \varphi_0(x) = W(x,x') \partial_0 \varphi_0(x)\,.
\label{Feyn:D0W}
\end{equation}
Properties of Wilson lines were investigated in many papers.
Many results now considered classics of HQET
were derived in the course of these studies
before HQET was invented in 1990.
In particular, the HQET Lagrangian~(\ref{HEET:spin0})
has been introduced as a technical device
for investigation of Wilson lines.

The lowest-energy state (``vacuum'') in HEET is a single electron at rest,
and it is convenient to use its energy as the zero level.
In the full theory, its energy is $M$, and
\begin{equation}
E = M + \varepsilon\,,
\label{Feyn:E}
\end{equation}
where $E$ is the energy of some state
(containing a single electron) in the full theory,
and $\varepsilon$ is its energy in HEET
(it is called the residual energy).
We can re-write this relation in a relativistic form:
\begin{equation}
P^\mu = M v^\mu + p^\mu\,,
\label{Feyn:P}
\end{equation}
where $P^\mu$ is the 4-momentum of some state
(containing a single electron) in the full theory,
$p^\mu$ is its momentum in HEET (the residual momentum),
and $v^\mu$ is 4-velocity of a reference frame
in which the electron always stays approximately at rest.
In other words, HEET is applicable if there exists such a 4-velocity $v$
that, after decomposition~(\ref{Feyn:P}),
the components of the electron residual momentum $p$ are always small,
and components of all photon momenta $p_i$ are also small:
\begin{equation}
p^\mu \ll M\,,\quad
p_i^\mu \ll M\,.
\label{Feyn:appl}
\end{equation}
This condition does not fix $v$ uniquely;
it can be varied by $\delta v \sim p/M$.
Effective theories corresponding to different choices of $v$
must produce identical physical predictions.
This requirement is called reparametrization invariance~\cite{LM:92}.
It produces relations between some quantities
of different orders in $1/M$.

We can re-write the Lagrangian~(\ref{Feyn:L})
in a relativistic form~\cite{G:90}:
\begin{equation}
L = \varphi_0^* i v \cdot D \varphi_0
+ (\text{light fields})\,.
\label{Feyn:Lv}
\end{equation}
This Lagrangian is not Lorentz-invariant,
because it contains a fixed vector $v$.
It gives the free propagator
\begin{equation}
S_0(p) = \frac{1}{p\cdot v + i0}\,.
\label{Feyn:S}
\end{equation}
The mass shell of the static electron is
\begin{equation}
p\cdot v = 0\,.
\label{Feyn:mshell}
\end{equation}
If we want to consider the spin-$\frac{1}{2}$ electron,
it is described by the 4-component (Dirac) spinor field $h_v$
which satisfies the condition
\begin{equation}
\rlap/v h_v = h_v
\label{Feyn:hv}
\end{equation}
(so that in the $v$ rest frame the field has only 2 upper components
non-vanishing).
The Lagrangian~\cite{G:90}
\begin{equation}
L = \bar{h}_{v0} i v\cdot D h_{v0}
+ (\text{light fields})
\label{Feyn:Lhv}
\end{equation}
gives the propagator
\begin{equation}
S_0(p) = \frac{1 + \rlap/v}{2}\,\frac{1}{p\cdot v + i0}
\label{Feyn:Sh}
\end{equation}
and the vertex $i e_0 v^\mu$~(\ref{Feyn:vert}).

And what can our friends from Qedland say about this theory?
They are not surprised.
The finite-mass free electron propagator $S_0(P)$
with $P=Mv+p$~(\ref{Feyn:P}), $M\to\infty$ can be approximated as
\begin{equation}
S_0(Mv+p) =
\frac{M + M\rlap/v + \rlap/p}{(Mv+p)^2-M^2+i0} =
\frac{1+\rlap/v}{2}\,\frac{1}{p\cdot v+i0}
+ \mathcal{O}\left(\frac{p}{M}\right)\,.
\label{Feyn:SP}
\end{equation}
Diagrammatically, it is related to the HEET propagator:
\begin{equation}
\raisebox{-3.5mm}{\begin{picture}(22,6.5)
\put(11,4.5){\makebox(0,0){\includegraphics{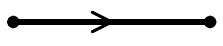}}}
\put(11,0){\makebox(0,0)[b]{$Mv+p$}}
\end{picture}} =
\raisebox{-3.5mm}{\begin{picture}(22,6.5)
\put(11,4.5){\makebox(0,0){\includegraphics{fr2.pdf}}}
\put(11,0){\makebox(0,0)[b]{$p$}}
\end{picture}}
+ \mathcal{O}\left(\frac{p}{M}\right)\,.
\label{Feyn:QED}
\end{equation}
When a QED vertex $i e_0 \gamma^\mu$ is sandwiched
between two propagators~(\ref{Feyn:Sh}),
it can be replaced by the HEET vertex $i e_0 v^\mu$:
\begin{equation}
\frac{1+\rlap/v}{2} \gamma^\mu \frac{1+\rlap/v}{2} =
\frac{1+\rlap/v}{2} v^\mu \frac{1+\rlap/v}{2}\,.
\label{Feyn:gamma}
\end{equation}
But what if there is an external spinor $u(P)$ after the vertex
(or $\bar{u}(P)$ before it)?
From the Dirac equation we have
\begin{equation*}
\rlap/v u(Mv+p) = u(Mv+p) + \mathcal{O}\left(\frac{p}{M}\right)\,,
\end{equation*}
so that we may insert the projectors $(1+\rlap/v)/2$
before $u(P_i)$ and after $\bar{u}(P_i)$, too,
and the replacement~(\ref{Feyn:gamma}) is applicable.
We have derived the HEET Feynman rules from the QED ones
in the limit $M\to\infty$.
Therefore, we again arrive at the HEET Lagrangian~(\ref{Feyn:Lhv})
which corresponds to these Feynman rules.

We have thus proved that at the tree level any QED diagram
is equal to the corresponding HEET diagram up to
$\mathcal{O}(p/M)$ corrections.
This is not true at loops,
because loop momenta can be arbitrarily large.
Renormalization properties of HEET
(anomalous dimensions, etc.) differ from those in QED.
QED loop diagrams can be decomposed into integration regions (Sect.~\ref{S:Reg}),
with some loops hard (momenta $\sim M$)
and some soft (momenta $\sim p$).
Then hard loops produce local interactions
(in the effective theory language,
they follow from local operators in the HEET Lagrangian);
soft loops can be calculated as in HEET.

\subsection{The heavy propagator and current}
\label{S:Prop}

In fact, the static electron propagator can be calculated exactly~\cite{YFS:61}!
Suppose we calculate the one-loop correction to the static electron propagator
in coordinate space.
Let us multiply this correction by itself.
We obtain an integral in $t_1$, $t_2$, $t_1'$, $t_2'$
with $0<t_1<t_2<t$, $0<t_1'<t_2'<t$.
The ordering of primed and non-primed integration times can be arbitrary.
The integration area is subdivided into six regions,
corresponding to the six diagrams:
\begin{equation}
\begin{split}
&\raisebox{-6.25mm}{\begin{picture}(23,14.5)
\put(11.5,9.875){\makebox(0,0){\includegraphics{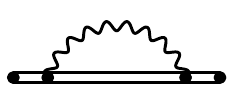}}}
\put(1,4.5){\makebox(0,0){$0$}}
\put(22,4.5){\makebox(0,0){$t$}}
\put(4.5,4.1){\makebox(0,0){$t_1$}}
\put(18.5,4.1){\makebox(0,0){$t_2$}}
\end{picture}}
\times
\raisebox{-6.25mm}{\begin{picture}(23,14.5)
\put(11.5,4.625){\makebox(0,0){\includegraphics{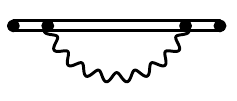}}}
\put(1,10.3){\makebox(0,0){$0$}}
\put(22,10.3){\makebox(0,0){$t$}}
\put(4.5,9.9){\makebox(0,0){$t_1'$}}
\put(18.5,9.9){\makebox(0,0){$t_2'$}}
\end{picture}}\\
&{}= \raisebox{-6.75mm}{\includegraphics{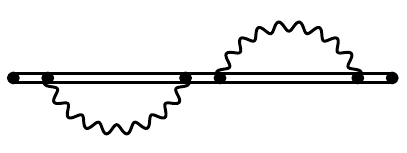}}
+ \raisebox{-6.75mm}{\includegraphics{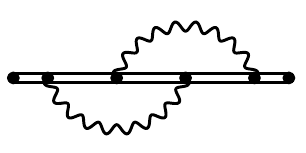}}
+ \raisebox{-6.75mm}{\includegraphics{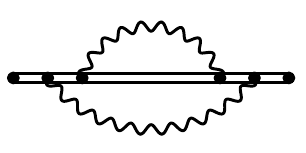}}\\
&{}+ \raisebox{-6.75mm}{\includegraphics{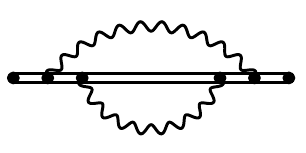}}
+ \raisebox{-6.75mm}{\includegraphics{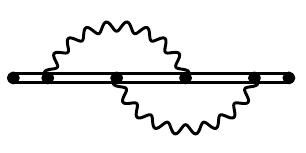}}
+ \raisebox{-6.75mm}{\includegraphics{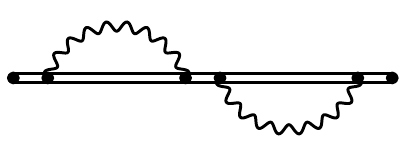}}
\end{split}
\label{Prop:exp}
\end{equation}
This is twice the 2-loop correction to the propagator.
Continuing this drawing exercise,
we see that the one-loop correction cubed
is $3!$ times the 3-loop correction, and so on.
Therefore, the exact all-order propagator
is the exponential of the one-loop correction:
\begin{equation}
S(t) = S_0(t) \exp w_1\,,\quad
w_1 = - \frac{e_0^2}{(4\pi)^{d/2}}
\left(\frac{it}{2}\right)^{2\varepsilon} \Gamma(-\varepsilon)
\left(\xi + \frac{2}{d-3}\right)
\label{Prop:S}
\end{equation}
($\xi = 1-a_0$).
In particular, in the $d$-dimensional Yennie gauge~\cite{FY:58}
\begin{equation}
a_0 = \frac{2}{d-3} + 1
\label{Prop:Yennie}
\end{equation}
the exact propagator~(\ref{Prop:S}) is free.

There are no corrections to the photon propagator in HEET~(\ref{Feyn:L})
because static-electron loops don't exist%
\footnote{This argument works up to the order $1/M^3$.
At $1/M^4$ a 4-photon interaction appears, see Sect.~\ref{S:Photonia}.
However, the only correction to the photon propagator
at this order vanishes~(\ref{Photonia:Pi0}).
The first non-vanishing correction involves two 4-photon vertices,
and appears at $1/M^8$.}.
Therefore, the photon field renormalization constant $Z_A=1$,
and hence $a = a_0$, $e = e_0$.
The renormalization constant of the static electron field $h$ and its anomalous dimension
are thus known exactly:
\begin{equation}
Z_h = \exp\left[- (a-3) \frac{\alpha}{4\pi\varepsilon}\right]\,,\quad
\gamma_h = 2 (a-3) \frac{\alpha}{4\pi}\,.
\label{Prop:gammah}
\end{equation}

\begin{figure}[ht]
\begin{center}
\begin{picture}(42,22)
\put(21,11){\makebox(0,0){\includegraphics{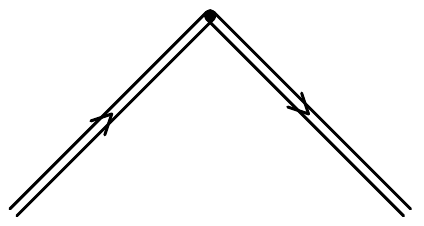}}}
\put(5,7){\makebox(0,0)[b]{$v$}}
\put(37,7){\makebox(0,0)[b]{$v'$}}
\end{picture}
\end{center}
\caption{Heavy--heavy current}
\label{F:HH0}
\end{figure}

Suppose the electron substantially changes its 4-velocity
(due to some hard-photon interaction).
In the HEET framework this can be described by the current (Fig.~\ref{F:HH0})
\begin{equation}
J_0 = \varphi_{v'0}^* \varphi_{\vphantom{v'}v0} = Z_J(\vartheta) J(\mu)\,,
\label{HH:J0}
\end{equation}
where $\cosh\vartheta = v\cdot v'$.
Its anomalous dimension
\begin{equation}
\Gamma(\vartheta) = \frac{d\log Z_J}{d\log\mu}
\label{HH:Gamma}
\end{equation}
is called the cusp anomalous dimension.
If $v'=v$ ($\vartheta = 0$) then $Z_J(0) = 1$ and $\Gamma(0) = 0$.
Exponentiation works for Wilson lines of any shape,
in particular, lines with a cusp.
Therefore the 1-loop formula for $\Gamma(\vartheta)$ is exact.

There are many methods to calculate $\Gamma(\vartheta)$.
The simplest one is based on unitarity.
The electron after the kick either remains itself
(probability $|F|^2$, where $F$ is its form factor)
or emits one or several protons:
\begin{equation*}
|F|^2 + \sum_{n=1}^\infty w_n = 1\,.
\end{equation*}
Up to the first order in $\alpha$,
\begin{equation*}
\Biggl|\raisebox{-4.8mm}{\includegraphics[scale=0.5]{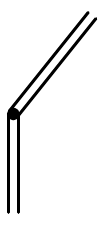}}
+ \raisebox{-4.8mm}{\includegraphics[scale=0.5]{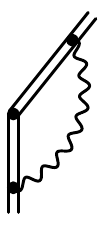}}\Biggr|^2
+ \int \Biggl|\raisebox{-4.8mm}{\includegraphics[scale=0.5]{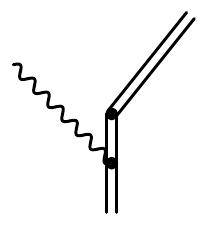}}
+ \raisebox{-4.8mm}{\includegraphics[scale=0.5]{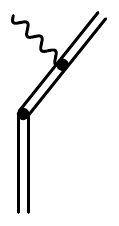}}\Biggr|^2
= 1\,.
\end{equation*}
In classical electrodynamics the spectrum of emitted radiation is~\cite{LL}
\begin{equation*}
dE = \frac{e^2}{2 \pi^2} (\vartheta \coth\vartheta - 1)\,d\omega\,,
\end{equation*}
and hence the probability of photon emission is
\begin{equation*}
dw = \frac{e^2}{2 \pi^2} (\vartheta \coth\vartheta - 1)
\frac{d\omega}{\omega}\,.
\end{equation*}
In dimensional regularization, by dimensions counting, this probability becomes
\begin{equation*}
dw = \frac{e^2}{2 \pi^2} (\vartheta \coth\vartheta - 1)
\frac{d\omega}{\omega^{1+2\varepsilon}}
\end{equation*}
(up to a factor which tends to 1 at $\varepsilon\to0$;
we don't need such a factor).
Hence the form factor is
\begin{equation*}
F = 1 - \frac{1}{2} \int_\lambda^\infty
\frac{e^2}{2 \pi^2} (\vartheta \coth\vartheta - 1)
\frac{d\omega}{\omega^{1+2\varepsilon}}\,,
\end{equation*}
where $\lambda$ is an infrared cutoff.
We need only the ultraviolet $1/\varepsilon$ divergence of $F$,
and so we have to introduce such a cutoff.
We obtain
\begin{equation*}
Z_J = 1 - 2 \frac{\alpha}{4\pi\varepsilon} (\vartheta \coth\vartheta - 1)\,,
\end{equation*}
and
\begin{equation}
\Gamma = 4 \frac{\alpha}{4\pi} (\vartheta \coth\vartheta - 1)\,.
\label{Prop:Gamma}
\end{equation}
It is given by the soft radiation function in classical electrodynamics,
and should be included in The Guinness Book of Records
as the anomalous dimension known for the longest time
(probably, $> 100$ years).

\subsection{Hadrons with a heavy quark}
\label{S:HQET}

The $B$ meson is the hydrogen atom of quantum chromodynamics,
the simplest non-trivial hadron.
In the leading approximation, the $b$ quark in it
just sits at rest at the origin and creates a chromoelectric field.
Light constituents (gluons, light quarks and antiquarks)
move in this external field.
Their motion is relativistic;
the number of gluons and light quark--antiquark pairs in this light cloud
is undetermined and varying.
Therefore, there are no reasons to expect that a non-relativistic
potential quark model would describe the $B$ meson well enough
(in contrast to the $\varUpsilon$ meson, where the non-relativistic
two-particle picture gives a good starting point).

Similarly, the $\Lambda_b$ baryon can be called the helium atom of QCD.
Unlike in atomic physics, where the hydrogen atom is much
simpler than helium,
the $B$ and $\Lambda_b$ are equally difficult.
Both have a light cloud with a variable number of relativistic particles.
The size of this cloud is the confinement radius $1/\Lambda_{\text{QCD}}$;
its properties are determined by large-distance nonperturbative QCD.

The analogy with atomic physics can tell us a lot about hadrons
with a heavy quark.
The usual hydrogen and tritium have identical chemical properties,
despite the fact that the tritium nucleus is three times heavier than the proton.
Both nuclei create identical electric fields, and both stay at rest.
Similarly, the $D$ and $B$ mesons have identical ``hadro-chemical'' properties,
despite the fact that the $b$ quark is three times heavier than the $c$.

The proton magnetic moment is of the order of the nuclear magneton $e/(2M_p)$,
and is much smaller than the electron magnetic moment $e/(2m_e)$.
Therefore, the energy difference between the states of the hydrogen atom
with total spins 0 and 1 (hyperfine splitting) is small
(of the order $m_e/M_p$ times the fine structure).
Similarly, the $b$-quark chromomagnetic moment is proportional to $1/M_b$
by dimensionality, and the hyperfine splitting between the $B$ and $B^*$ mesons
is small (proportional to $1/M_b$).
Unlike in atomic physics, both ``gross''-structure intervals
and fine-structure intervals are just some numbers times $\Lambda_{\text{QCD}}$,
because the light components are relativistic
(the practical success of constituent quark models shows that
these dimensionless numbers for fine splittings can be rather small,
but they contain no small parameter).

In the limit $M\to\infty$,
the heavy-quark spin does not interact with the gluon field.
Therefore, it may be rotated at will, without changing the physics.
Such rotations can transform the $B$ and $B^*$ into each other;
they are degenerate and have identical properties in this limit.
This heavy-quark spin symmetry yields many useful relations
among heavy-hadron form factors.
Not only the orientation, but also the magnitude of the heavy-quark spin
is irrelevant in the infinite-mass limit. 
We can switch off the heavy-quark spin, making it spinless,
without affecting the physics.
This leads to a supersymmetry group called
the superflavor symmetry.
It can be used to predict properties of hadrons containing
a scalar or vector heavy quark.
Such quarks exist in some extensions of the Standard Model
(for example, supersymmetric or composite extensions).

This idea can also be applied to baryons with two heavy quarks.
They form a small-size bound state
(with a radius of order $1/(M \alpha_s)$)
which has spin 0 or 1 and is antitriplet in color.
Therefore, these baryons are similar to mesons with a heavy antiquark
that has spin 0 or 1.
The accuracy of this picture cannot be high, however,
because even the radius of the $bb$ diquark is only a few times
smaller than the confinement radius.

Let us consider mesons with the quark contents $\bar{Q}q$,
where $Q$ is a heavy quark with mass $M$ ($b$ or $c$),
and $q$ is a light quark ($u$, $d$, or $s$).
As discussed above, the heavy-quark spin is inessential
in the limit $M\to\infty$, and may be switched off.
In a world with a scalar heavy antiquark,
$S$-wave mesons have angular momentum and parity $j^P=\frac{1}{2}^+$;
$P$-wave mesons have $j^P=\frac{1}{2}^-$ and $\frac{3}{3}^-$.
The energy difference between these two $P$-wave states (fine splitting)
is a constant times $\Lambda_{\text{QCD}}$ at $M\to\infty$,
just like the splittings between these $P$-wave states and the ground state;
however, this constant is likely to be small.

In our real world, the heavy antiquark $\bar{Q}$
has spin and parity $s_Q^P=\frac{1}{2}^-$.
The quantum numbers in the above paragraph are those
of the cloud of light fields of a meson.
Adding the heavy-antiquark spin, we obtain, in the limit $M\to\infty$,
a degenerate doublet of $S$-wave mesons with spin and parity
$s^P=0^-$ and $1^-$,
and two degenerate doublets of $P$-wave mesons,
one with $s^P=0^+$ and $1^+$, and the other with $s^P=1^+$ and $2^+$.
At a large but finite heavy-quark mass $M$,
these doublets are not exactly degenerate.
Hyperfine splittings, equal to to some dimensionless numbers
times $\Lambda_{\text{QCD}}^2/M$, appear.
It is natural to expect that hyperfine splittings in $P$-wave mesons
are less than in the ground-state $S$-wave doublet,
because the characteristic distance between the quarks is larger
in the $P$-wave case.
Note that the $1^+$ mesons from the different doublets
do not differ from each other by any exactly conserved quantum numbers,
and hence can mix.
They differ by the angular momenta of the light fields,
which is conserved up to $1/M$ corrections;
therefore, the mixing angle should be
of the order of $\Lambda_{\text{QCD}}/M$.

Mesons with $q=u$ and $d$ form isodoublets;
together with isosinglets with $q=s$, they form $SU(3)$ triplets.

\begin{figure}[ht]
\begin{center}
\begin{picture}(60,72)
\put(27,46){\makebox(0,0){\includegraphics{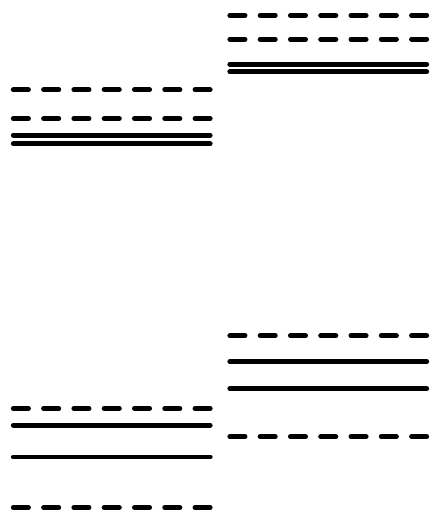}}}
\put(3,26.13457){\makebox(0,0){$0^-$}}
\put(3,29.35565){\makebox(0,0){$1^-$}}
\put(3,56.944){\makebox(0,0){$1^+$}}
\put(3,59.8238){\makebox(0,0){$2^+$}}
\put(51,33.0729){\makebox(0,0){$0^-$}}
\put(51,35.8171){\makebox(0,0){$1^-$}}
\put(51,64.2603){\makebox(0,0){$1^+$}}
\put(51,67.0546){\makebox(0,0){$2^+$}}
\put(16,15){\makebox(0,0){$B^+ = \bar{b}u$}}
\put(16,10){\makebox(0,0){$B^0 = \bar{b}d$}}
\put(38,15){\makebox(0,0){$B_s^0 = \bar{b}s$}}
\put(16,0){\makebox(0,0){$\bar{D}{}^- = \bar{c}d$}}
\put(16,5){\makebox(0,0){$\bar{D}{}^0 = \bar{c}u$}}
\put(38,5){\makebox(0,0){$\bar{D}{}_s^- = \bar{c}s$}}
\put(16,24.13457){\makebox(0,0){$B$}}
\put(16,31.35565){\makebox(0,0){$B^*$}}
\put(16,55.94389){\makebox(0,0){$B_1$}}
\put(16,60.82378){\makebox(0,0){$B_2^*$}}
\end{picture}
\end{center}
\caption{Mesons with $\bar{b}$ (solid lines) and $\bar{c}$ (dashed lines).}
\label{F:Mesons}
\end{figure}
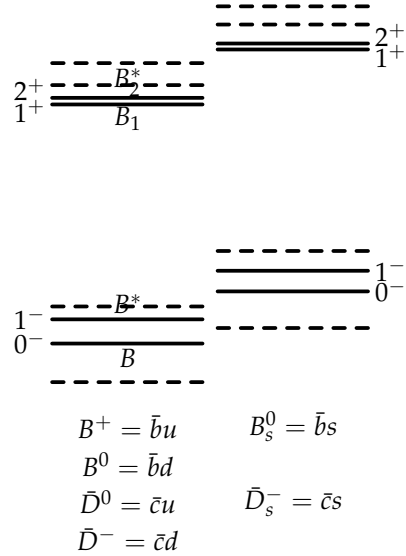

The experimentally observed mesons containing
the $\bar{b}$ antiquark are shown in Fig.~\ref{F:Mesons}.
The mesons $B_1$ and $B_2^*$ form a doublet,
with the quantum numbers of the light fields $j^P=\frac{3}{2}^+$.
The second $P$-wave doublet with $j^P=\frac{1}{2}^+$ is suspiciously absent.
The $j^P=\frac{3}{2}^+$ mesons decays into the ground-state $\frac{1}{2}^-$ doublet
and a $\pi$ meson in $D$-wave ($l=2$);
the pion momentum is small, these decays are strongly suppressed,
and the $j^P=\frac{3}{2}^+$ mesons are narrow.
The $j^P=\frac{1}{2}^+$ mesons decays into the ground-state ones
and a $\pi$ meson in $S$-wave ($l=0$);
these decays are not suppressed, and hence the $j^P=\frac{1}{2}^+$ mesons
are very wide.
Hyperfine splitting of the $P$-wave doublet is smaller that that of the $S$-wave one.

In the leading approximation,
the spectrum of $\bar{c}$-containing mesons
is obtained from the spectrum of $\bar{b}$-containing mesons
simply by a shift by $M_c-M_b$.
The spectrum of mesons containing $\bar{c}$
is also shown in Fig.~\ref{F:Mesons}.
It is positioned in such a way that the weighted average energies
of the ground-state doublets $(M_B + 3 M_{B^*})/4$ and $(M_D + 3 M_{D^*})/4$ coincide.
Hyperfine splittings are smaller for $B$ mesons than for $D$ mesons, as expected.

In $S$-wave $Qqq$ baryons,
the light-quark spins can add to give $j^P=0^+$ or $1^+$.
In the first case their spin wave function is antisymmetric;
the Fermi statistics and the antisymmetry in color
require an antisymmetric flavor wave function.
Hence the light quarks must be different;
if they are $u$, $d$, then their isospin is $I=0$.
With the heavy-quark spin switched off,
this gives the $0^+$ baryon $\Lambda_Q$ with $I=0$.
If one of the light quarks is $s$,
we have the isodoublet $\Xi_Q$,
which forms an $SU(3)$ antitriplet together with $\Lambda_Q$.
With the heavy-quark spin switched on,
these baryons have $s^P=\frac{1}{2}^+$.
In the $1^+$ case, the flavor wave function is symmetric.
If the light quarks are $u$, $d$, then their isospin is $I=1$.
This gives the $1^+$ isotriplet $\Sigma_Q$;
with one $s$ quark, we obtain the isodoublet $\Xi'_Q$;
and with two $s$ quarks, the isosinglet $\Omega_Q$.
Together, they form an $SU(3)$ sextet.
With the heavy-quark spin switched on,
we obtain the degenerate doublets with $s^P=\frac{1}{2}^+$, $\frac{3}{2}^+$:
$\Sigma_Q$, $\Sigma^*_Q$; $\Xi'_Q$, $\Xi^*_Q$; $\Omega_Q$, $\Omega^*_Q$.
The hyperfine splittings in these doublets
are of the order of $\Lambda_{\text{QCD}}^2/M$.
Mixing between $\Xi_Q$ and $\Xi'_Q$ is suppressed
both by $1/M$ and by $SU(3)$.

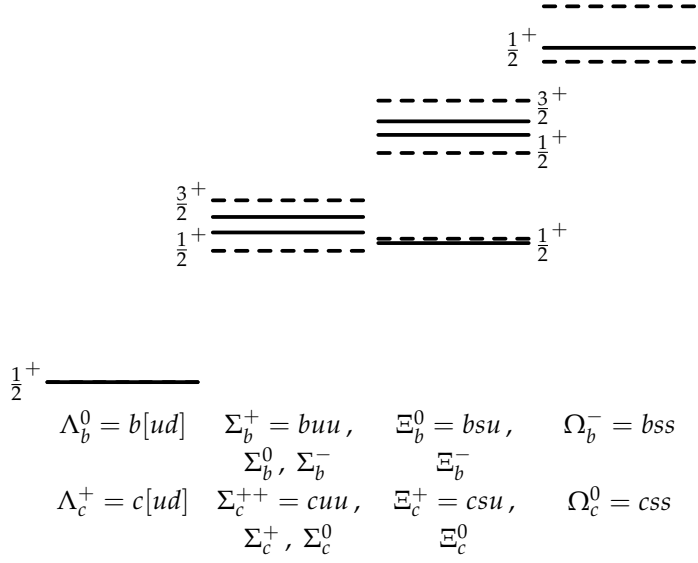
\begin{figure}[ht]
\begin{center}
\begin{picture}(100,72)
\put(49,46){\makebox(0,0){\includegraphics{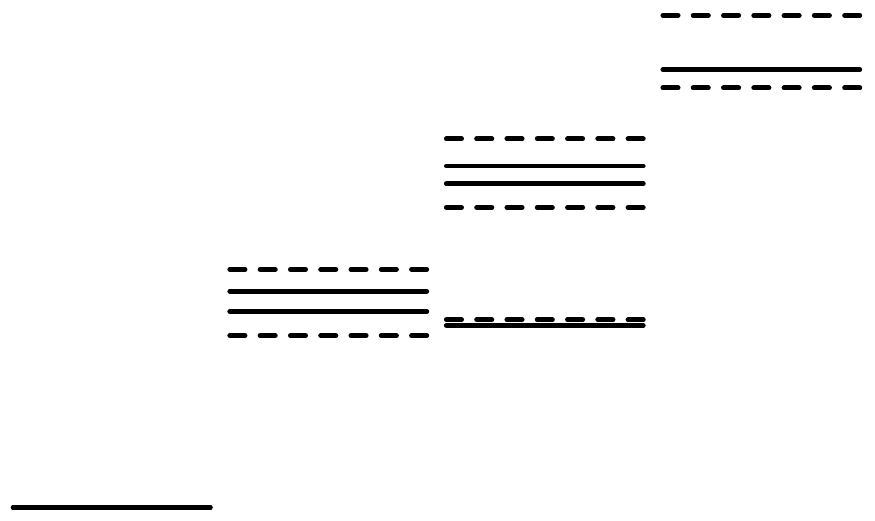}}}
\put(3,21){\makebox(0,0){$\frac{1}{2}^+$}}
\put(25,38.9149){\makebox(0,0){$\frac{1}{2}^+$}}
\put(25,44.9756){\makebox(0,0){$\frac{3}{2}^+$}}
\put(73,39.5008){\makebox(0,0){$\frac{1}{2}^+$}}
\put(73,51.8946){\makebox(0,0){$\frac{1}{2}^+$}}
\put(73,57.6967){\makebox(0,0){$\frac{3}{2}^+$}}
\put(69,65.479){\makebox(0,0){$\frac{1}{2}^+$}}
\put(16,15){\makebox(0,0){$\Lambda_b^0 = b[ud]$}}
\put(38,15){\makebox(0,0){$\Sigma_b^+ = buu\,,$}}
\put(38,10){\makebox(0,0){$\Sigma_b^0\,,\;\Sigma_b^-$}}
\put(60,15){\makebox(0,0){$\Xi_b^0 = bsu\,,$}}
\put(60,10){\makebox(0,0){$\Xi_b^-$}}
\put(82,15){\makebox(0,0){$\Omega_b^- = bss$}}
\put(16,5){\makebox(0,0){$\Lambda_c^+ = c[ud]$}}
\put(38,5){\makebox(0,0){$\Sigma_c^{++} = cuu\,,$}}
\put(38,0){\makebox(0,0){$\Sigma_c^+\,,\;\Sigma_c^0$}}
\put(60,5){\makebox(0,0){$\Xi_c^+ = csu\,,$}}
\put(60,0){\makebox(0,0){$\Xi_c^0$}}
\put(82,5){\makebox(0,0){$\Omega_c^0 = css$}}
\end{picture}
\end{center}
\caption{Baryons with $b$ (solid lines) and $c$ (dashed lines).}
\label{F:Baryons}
\end{figure}

The experimentally observed baryons containing $b$
are shown in Fig.~\ref{F:Baryons}.
In the third column, the lowest state $\Xi_b$
is followed by the doublet $\Xi'_b$, $\Xi^*_b$.
The $\Omega^*_b$ baryon has not yet been observed.
The spectrum of baryons containing $c$
is also shown.
It is positioned in such a way that the ground-state baryons
$\Lambda_b$ and $\Lambda_c$ coincide.

In the leading $M_b \to \infty$ approximation,
the masses $M_B$ and $M_{B^*}$ are both equal to $M_b + \bar{\Lambda}$,
where $\bar{\Lambda}$ is the energy of the ground state
of the light fields in the chromoelectric field of the $\bar{b}$ antiquark.
This energy $\bar{\Lambda}$ is of the order of $\Lambda_{\text{QCD}}$.
The excited states of the light fields have energies $\bar{\Lambda}_i$,
giving excited degenerate doublets with masses $M_b + \bar{\Lambda}_i$.

There are two $1/M_b$ corrections to the masses.
First, the $\bar{b}$ antiquark has an average momentum squared $\mu_\pi^2$,
which is of order of $\Lambda_{\text{QCD}}^2$.
Therefore, it has a kinetic energy $\mu_\pi^2/(2 M_b)$.
Second, the $\bar{b}$ chromomagnetic moment interacts with the chromomagnetic
field created by light constituents at the origin, where the $\bar{b}$ stays.
This chromomagnetic field is proportional to
the angular momentum of the light fields $\vec{\jmath}$.
Therefore, the chromomagnetic interaction energy is proportional to
\begin{equation*}
\vec{s}_{Q}\cdot\vec{\jmath} =
\frac{1}{2} \left[s(s+1) - s_{Q}(s_{Q}+1) - j(j+1)\right] =
\left\{
\begin{array}{ll}
\displaystyle         - \frac{3}{4}\,, & s=0\,, \\[2mm]
\displaystyle\phantom{-}\frac{1}{4}\,, & s=1\,,
\end{array}
\right.
\end{equation*}
where $\vec{s}=\vec{s}_Q+\vec{\jmath}$ is the meson spin.
If we denote this energy for $B$ by $-\mu_G^2/(2 M_b)$,
then for $B^*$ it will be $(1/3)\mu_G^2/(2 M_b)$.
Here $\mu_G^2$ is of order of $\Lambda_{\text{QCD}}^2$.
The $B$, $B^*$ meson masses with $1/M_b$ corrections taken into account
are given by the formulae
\begin{equation}
\begin{split}
&M_B = M_b + \bar{\Lambda} + \frac{\mu_\pi^2 - \mu_G^2}{2 M_b}
+ \mathcal{O}\left(\frac{\Lambda_{\text{QCD}}^3}{M_b^2}\right)\,,\\
&M_{B^*} = M_b + \bar{\Lambda} + \frac{\mu_\pi^2 + \frac{1}{3} \mu_G^2}{2 M_b}
+ \mathcal{O}\left(\frac{\Lambda_{\text{QCD}}^3}{M_b^2}\right)\,.
\end{split}
\label{Hadron:mass}
\end{equation}
The hyperfine splitting is
\begin{equation*}
M_{B^*} - M_B = \frac{2 \mu_G^2}{3 M_b}
+ \mathcal{O}\left(\frac{\Lambda_{\text{QCD}}^3}{M_b^2}\right)\,.
\end{equation*}
Taking into account $M_{B^*} + M_B = 2 M_b + \mathcal{O}(\Lambda_{\text{QCD}})$,
we obtain
\begin{equation*}
M_{B^*}^2 - M_B^2 = \frac{4}{3} \mu_G^2
+ \mathcal{O}\left(\frac{\Lambda_{\text{QCD}}^3}{M_b}\right)\,.
\end{equation*}
The difference $M_{D^*}^2 - M_D^2$ is given by a similar formula,
with $M_c$ instead of $M_b$.
Therefore, the ratio
\begin{equation}
\frac{M_{B^*}^2 - M_B^2}{M_{D^*}^2 - M_D^2} = 1
+ \mathcal{O}\left(\frac{\Lambda_{\text{QCD}}}{M_{c,b}}\right)\,.
\label{Hadron:ratio}
\end{equation}
Experimentally, this ratio is 0.88.
This is a spectacular confirmation of the idea
that violations of the heavy-quark spin symmetry
are proportional to $1/M$.

Let us now discuss the $B$-meson leptonic decay constant $f_B$.
It is defined by
\begin{equation*}
\bigl<0|\bar{b}\gamma^\mu\gamma_5u|B(p)\bigr> = i f_B p^\mu\,,
\end{equation*}
where the one-particle state is normalized in the usual Lorentz-invariant way:
\begin{equation*}
\bigl<B(p')|B(p)\bigr> = 2 p^0 (2\pi)^3 \delta(\vec{p}'-\vec{p})\,.
\end{equation*}
This relativistic normalization becomes nonsensical in the limit $M_b \to \infty$,
and in that case the non-relativistic normalization
\begin{equation*}
\strut_{\text{nr}\!}\bigl<B(p')|B(p)\bigr>_{\!\text{nr}} = (2\pi)^3 \delta(\vec{p}'-\vec{p})
\end{equation*}
should be used instead.
Then, for the $B$ meson at rest,
\begin{equation*}
\bigl<0|\bar{b}\gamma^0\gamma_5u|B\bigr>_{\!\text{nr}} = \frac{i M_B f_B}{\sqrt{2 M_B}}\,.
\end{equation*}
Denoting this matrix element (which is mass-independent at $M_b \to \infty$)
by $i F/\sqrt{2}$, we obtain
\begin{equation}
f_B = \frac{F}{\sqrt{M_b}} \left[1 +
\mathcal{O}\left(\frac{\Lambda_{\text{QCD}}}{M_b}\right)\right]\,,
\label{Hadron:fB}
\end{equation}
and hence
\begin{equation}
\frac{f_B}{f_D} = \sqrt{\frac{M_c}{M_b}} \left[1 +
\mathcal{O}\left(\frac{\Lambda_{\text{QCD}}}{M_{c,b}}\right)\right]\,.
\label{Hadron:fRatio}
\end{equation}

\section{Conclusion}
\label{S:Conc}

In the past only renormalizable theories were considered well-defined:
they contain a finite number of parameters,
which can be extracted from a finite number of experimental results
and used to predict results of an infinite number of other potential measurements.
Non-renormalizable theories were rejected
because their renormalization at all orders in non-renormalizable interactions
involve infinitely many parameters,
so that such a theory has no predictive power.
This principle is absolutely correct,
if we are impudent enough to pretend that our theory
describes the Nature up to arbitrarily high energies
(or down to arbitrarily small distances).

At present we accept the fact that our theories only describe the Nature
at sufficiently low energies (or sufficiently large distances).
They are effective low-energy theories.
Such theories contain all operators (allowed by the relevant symmetries)
in their Lagrangians.
They are necessarily non-renormalizable.
This does not prevent us from obtaining definite predictions
at any fixed order in the expansion in $E/M$,
where $E$ is the characteristic energy
and $M$ is the scale of new physics.
Only if we are lucky and $M$ is many orders of magnitude larger
than the energies we are interested in,
we can neglect higher-dimensional operators in the Lagrangian
and work with a renormalizable theory.

We can add higher-dimensional contributions to the Lagrangian,
with further unknown coefficients.
To any finite order in $1/M$, the number of such coefficients is finite,
and the theory has predictive power.
For example, if we want to work at the order $1/M^4$,
then either a single $1/M^4$ (dimension 8) vertex
or two $1/M^2$ ones (dimension 6) can occur in a diagram.
UV divergences which appear in diagrams with two dimension 6 vertices
are compensated by renormalizing these 2 operators plus dimension 8 counterterms.
So, the theory can be renormalized.
The usual arguments about non-renormalizability
are based on considering diagrams with arbitrarily many vertices
of nonrenormalizable interactions (operators of dimensions $>4$);
this leads to infinitely many free parameters in the theory.

The Standard Model does not describe \textit{all} physics
up to \textit{infinitely high} energies
(or down to \textit{infinitely small} distances).
At least, quantum gravity becomes important at the Planck scale.
SM does not explain dark matter and baryon asymmetry of the Universe.
What can appear at some very short distance scale?
\begin{itemize}
\item supersymmetry;
\item compositeness;
\item extra dimensions;
\item non-pointlike objects: strings / superstrings / branes;
\item something we cannot imagine at present.
\end{itemize}
We can construct scenarios for new physics searches based on
some known variants of the next theory (more fundamental than SM).
I.\,e.\ we can investigate some finite number of directions of departure from SM,
but an infinite number of directions which we cannot now imagine remain unexplored
--- this is a set of measure 0.
What we need is a systematic model-independent approach
for searching of some \textit{absolutely unknown}
new physics at small distances.
It produces new local interactions of the SM fields.
This approach is called Standard Model Effective Theory (SMEFT),
see, e.\,g., \cite{Warsaw,SMEFT}.

\section*{Acknowledgements}

The work has been supported by the Russian Ministry of Science and Higher Education.

\end{document}